# CHARACTERIZATION OF LOAD BEARING METROLOGICAL PARAMETERS IN REPTILIAN EXUVIAE IN COMPARISON TO PRECISION FINISHED CYLINDER LINER SURFACES

H. A. Abdel-Aal [1*] M. El Mansori[2]

[1] *The University of North Carolina at Charlotte, 9201 University City Blvd, Charlotte, NC 28223-0001, USA*
**corresponding author: Hisham.abdelaal@UNCC.edu*
[2]*Ecole Nationale Supérieure d'Arts et Métiers, 2 Cours des Arts et Métiers, Aix En Provence Cedex 1, France*

## ABSTRACT

Design of precise functional surfaces is essential for many future applications. In the technological realm, the accumulated experience with the construction of such surfaces is not sufficient. Nature provides many examples of dynamic surfaces worthy of study and adoption, at least as a concept, within human engineering. In this work, we probe load-bearing features of the ventral skin of snake surfaces. We examine the structure of two snake species that mainly move by rectilinear locomotion. These are Python regius (pythonidae) and Bittis gabonica (Vipridae). To this end, we focus on the load bearing characteristics of the ventral skin surface (i.e. the $R_k$ family of parameters). Therefore, we draw detailed comparison between the reptilian surfaces and two sets of technological data. The first set pertains to an actual commercial cylinder liner, whereas, the second set is a summary of recommended surface finish metrological values for several commercial cylinder liner manufacturers. The results highlight several similarities between the two types of surfaces. In particular, we show that there is a striking correspondence between the sense of texture engineering within both surfaces (although that their construction evolved along entirely different paths). We also show that reptilian surfaces manifest a high degree of specialization with respect to habitat constraints on wear resistance and adhesive effects.

## Nomenclature

| | |
|---|---|
| AFM | Atomic Force Microscopy |
| AP | Anterior Posterior |
| LL-RL | Lateral axis |
| AFLC | Abbot Firestone Curve |
| BDP | Bottom Dead Point |
| G | Body Girth in Snakes |
| MES | Man Engineered Systems |
| MP | Middle Point |
| $R_{pk}$ | Reduced Peak Height |
| $R_{vk}$ | Reduced Valley Depth |
| $R_k$ | Core Roughness Depth |
| Sa | Surface roughness Average height |
| $S_q$ | |
| $S_{ku}$ | Surface Kurtosis |
| $S_{sk}$ | Surface Skewness |
| SEM | Scan Electron Microscopy |
| TDP | Top Dead Point |
| WLI | White Light Interferometer |

## 1. Introduction

Many next generation products must offer optimal performance. Therefore, application of ultra-precise, complex and structured surfaces, such as those textured through plateau honing for improved lubrication capability, is an active area of research. In seeking inspirations for such custom designed surfaces, many engineers turn toward natural systems (i.e., bio-species, plants, insects etc.) because of their unique features. These include superior functionality, the ability to harness functional complexity to achieve optimal performance, and harmony between shape form and function. From a tribology perspective, the existence of surfaces that are design features, of the particular species, intended to facilitate functional performance is a point of deserving interest.

Tribological investigations often deal with complex systems that, while nominally homogeneous, are practically compositionally heterogeneous. Compositional heterogeneity is either inherent (structural), or evolutionary (functional) [1]. Inherent heterogeneity is due to initial variation in composition, material selection, component chemistry, etc. Evolutionary heterogeneity, on the other hand, arises because of the evolution of the local response of different parts of a sliding assembly during operation. Subsystem components, for example, since they entertain different loads will react in a manner that is proportional to the local loading conditions. Distinct responses cause system subcomponents to evolve into entities that differ from their initial state. System heterogeneity, thus, introduces a level of functional complexity to the sliding assembly. Functional complexity, in turn, characterizes the interaction of system subcomponents, and of the system as a unit, with the surroundings. Most of such interaction, it is to be noted, takes place through the surface. Natural systems, regardless of the degree of functional complexity inherited within, display harmonious characteristics and an ability to self-regulate. Whence, as a rule they operate at an optimal state marked by economy-of-effort. Man-engineered systems (MES), in contrast, do not exhibit such a level of optimized performance.

Much of the ability of natural systems to self regulate is attributed to optimized relationship between shape, form and function especially when surface design is considered. That is, shape and form in natural systems contribute to optimal function. Such customization, however, is not advanced in MES. Bio-species, in that respect, offer many a lesson. This is because biological materials, through million years of existence, have evolved optimized topological features that enhance wear and friction resistance [2]. One species of remarkable tribological performance that may serve as an inspiration for optimal surface texturing is that of snakes. This is due to the Objective-targeted design features associated with their mode of legless locomotion.

Snakes lack legs and use the surface of the body itself to generate propulsion on the ground during locomotion. For such a purpose, frictional tractions are necessary, in order to transmit forces to the ground. Depending on the snake species, type of movement, environment and preferred substrate, different parts of the body must have different functional requirements and therefore different frictional properties. That is the snake species, is a true representative of a heterogeneous tribo-system with a high degree of functional complexity, despite which, they do not suffer damaging levels of wear and tear.

Many researchers investigated the intriguing features of the serpentine family. Adam and Grace [3] studied the ultra structure of pit organ epidermis in Boid snakes to understand infrared sensing mechanisms. Johnna et al [4] investigated the permeability of shed skin of pythons (python molurus, Elaphe obsolete) to determine the suitability as a human skin analogue. Mechanical behavior of snakeskin was also a subject of several studies as well. Jayne [5] examined the loading curves of six different species in uni-axial extension. His measurements revealed substantial variation in loads and maximum stiffness among samples from different dorso-ventral regions within an individual and among homologous samples from different species. Rivera et al.[6] measured the mechanical properties of the integument of the common garter snake (Thomnophis sirtalis-Serpentine Colubridae). They examined mechanical properties of the skin along the body axis. Data collected revealed significant differences in mechanical properties among regions of the body. In particular, and consistent with the demands of macrophagy, it was found that the pre-pyloric skin is more compliant than post pyloric skin. Prompted by needs to design bio-inspired robots several researchers probed the frictional features of snake motion to understand the mechanisms responsible for regulating legless locomotion. Hazel et al [7] used AFM scanning to probe the nano-scale design features of three snake species. The studies of Hazel and Grace revealed the asymmetric features of the skin ornamentation to which both authors attributed frictional anisotropy.

In order to mimic the beneficial performance features of the skin, an engineer should be provided with parametric guidelines to aid with the objective-oriented design process. These should not only be dimensional. Rather, they should extend to include metrological parameters used to characterize tribological performance of surfaces within the MES domain. Thus, in order to deduce design rules there is a need to quantify the relationship governing micro-structure, strength ,and topology of the bio-surface; exploring the quantitative regulation of macro and micro texture, and finally devising working formulae that describe (and potentially predict) load carrying capacity during locomotion in relation to geometrical configuration at both the micro and the macro scale.

One informative method to reveal the sense of design of a surface is to compare the bearing curve characteristics of that surface to comparative operating surfaces; and to correlate the textural values to it's operation conditions. This work exploits such a methodology to investigate the bearing characteristics of two biological surfaces. To this end, we investigate the load bearing curves of the ventral scales of a Python regius and a Bittis Gabonica. The two snakes share roughly the same snout-to-vent length, and move by a rectilinear locomotion. The Bittis species, however, is almost five times the body mass of the Python. The difference in mass reflects on the constructal details of body geometry including the topology of the ventral texture. As such, studying the bearing curve characteristics of both species provides an in-depth look into the contribution of texture construction in snakes to efficiency of locomotion (minimized frictional waste and structural integrity of surface). The main thrust of this work is the analogy between textural construction of the reptilian surfaces and that of precise finished technological surfaces. Each of the examined surfaces evolved along a separate conceptual path. As such, attention is devoted to probing the correspondence in design paradigms of both surfaces.

## 2. Background

The current study examined shed skin from two snake species: a Gabon Viper (Bitis gabonica Rhinoceros) and a Royal Python (Python regius). Table 1 presents a summary of species taxonomy and major dimensional features

*Table 1* summary of species taxonomy and major dimensional features

| | Bitis Gabonic | Python regius |
|---|---|---|
| *Family* | Viperidae | Pythonidae |
| *Subfamily* | Viperinae | Python |
| *Genus* | Bitis | P. regius |
| *Species* | B. gabonica | |
| *Length (cm)* | 150 | 150 |
| *Number of ventral scales* | 132 | 208 |
| *Ratio of length to maximum diameter* | 5.2 | 10.2 |
| *Mass (Kg)* | 8 | 1.3 |
| *Average area of ventral scale mm$^2$* | 424.75 | 102.35 |
| *Maximum length of fibrils ($\mu$m)* | 0.75 | 1 |
| *Maximum ventral scale aspect ratio* | 5.99 | 3.142 |
| *Minimum Ventral scale aspect ratio* | 4.96 | 1.75 |

#### 2.1.1. The species

*Bitis Gabonica* is a venomous viper species. It is the largest of the Bitis genus [8]. This species occupies a wide range of sub-Saharan Africa particularly in the west part. Its habitat is mainly concentrated in rain forests and woodlands in the vicinity. Two subspecies are currently recognized [8], Bitis gabonica gabonica (BGG) and Bitis Gabonica Rhinoceros (BGR) (the subject of this study). BGR tends to move slowly and due to anatomical features, it tends mostly to adopt a rectilinear mode of locomotion. The length of the snake averages between 120-160 cms, and may weigh up to 8.5 Kgs. As such, it is one of the heaviest species. The length of the head plus trunk (snout-to-vent length) is four to five times its maximum body circumference [9]. Figure 1 depicts the snake and the details of different parts of the body. The main figure shows the overall shape and appearance of the reptile. The head of the reptile is visibly large and is triangular. The snake also is heavy bodied with a large girth. The cross section of the body, parallel to the lateral axis appears to be elliptical. Figure 1-a details the structure of the skin at the forehead region. The figure depicts SEM pictures of a square area (250 μm by 250 μm) on the forehead of the reptile at a magnification of X=500. The structure of the skin manifests several quasi-circular interconnected scales. The area of an individual scale varies within the range $200\ \mu m^2 \leq A_{vs} \leq 300\ \mu m^2$.

Figure 1-b shows SEM pictures of a rectangular dorsal skin spot adjacent to the forehead. The microstructure manifests a ridge-dominated construction. The ridges appear to be micron sized skin folds with gaps in between forming micro-channels. Such a composition is more visible in Figure 1-c, which depicts the details of the same skin spot at a higher magnification (x=2000). Note that the orientation of the skin ridges is rather random. Despite the array of coloration within the dorsal skin, it appears that the essential structural details of the skin do not vary by color. Figure 1-d depicts the structural details of a rectangular light-colored skin spot located on the dorsal side of the reptile within the rear half the body. The SEM micrograph details the junction between two adjacent scales (with partial appearance of the hinge region). Again ridge like skin folds appear to be dominant within the cell microstructure.

The microstructure of the ventral scales of the reptile appears to be of a shingle overlapping style (see figures 3-e and 3-f). The edges of each shingle appear to be serrated (similar to the teeth of a saw). The serration comprises a series of pikes (denticulations). These are organized in a step like formation. The three dimensional arrangement of the denticulations is detailed in figures 3-i and 3-j, which depict AFM scans of a square areas of sides 50 μm and 10 μm, respectively.

Figure 1-j reveals that the denticulations form a step-like structure. Further, it reveals that the size and length of the denticulations is quasi uniform.

The BGR, species is distinctive not only by its massive triangular head, but also by the presence of two horn-like structures at the dorsal side of the snout. This particular region is depicted in figures 2-a, and 2-b. Figures 2-c and 2-d meanwhile depict the structural detail of the skin within the horn region. Figure 2-c depicts the junction between two adjacent skin cells (magnification x=100). Figure 2-d, meanwhile, presents the overall structural configuration within the horn region (actual location is the left hand side-left lateral- horn) at a magnification of x=1000. Again, similar to the forehead region, the cell structure is quasi circular with the diameter of each circular cell ranging between 20 $\mu$m $\leq$ d $\leq$ 30 $\mu$m.

*Python Regius* is a non-venomous python species native to Africa. It is the smallest of the African pythons and is popular in the pet trade. The build is stocky while the head is relatively small. The color pattern is typically black with light brown-green side and dorsal blotches. The belly is white or cream that may or may not include scattered black markings. Figure 3 depicts the general features of the skin. Dorsal skin comprises several dark and light colored blotches.

The ventral skin, on the other hand, is mainly cream in color with occasional black markings. Two SEM pictures, at different magnifications (x=1000 and x=10000), are provided for each skin color. As shown, scale micro-features comprise fibrils arranged in rows. The shape of the fibrils, and the spacing between fibril waves, appear to be inconsistent. They vary by location and by color of skin. Fibril tips point toward the posterior end of the reptile. The shape of fibril tips varies with color of skin. In the dark-colored dorsal scales, fibrils are tapered and have a sharp tip. Scales within the bright colored and the ventral regions comprise rounded-tip fibrils of uniform width throughout the fibril length. Moreover, the density of the fibrils seems to be differ by region color (denser within the dark colored region).

The build of the snake is non-uniform. The head-neck region as well as that of the tail, is thinner than the-region containing the trunk. The trunk meanwhile is the region of the body where most of the snake body mass is concentrated. It is more compact and thicker than other parts. Consequently, most of the load bearing upon sliding occurs within the trunk region. The tail section also is rather conical in shape, although considerably thinner than the trunk. The overall cross section of the body is quasi elliptical rather than circular (the circumference of the upper half of the cross section is lengthier than the circumference of the lower half)

The general shape of the denticulations (fibrils) in the Python differs according to the color of the skin rather than by location (dorsal vs. ventral). Figure 4 (a and b) shows such a difference. The figure depicts AFM scans of a square area (10 $\mu$m by 10 $\mu$m) on the ventral side (Figure 4-a) and the dorsal dark spots (DDS) (Figure 4-b) of the Python. The general shape of the DDS is essentially triangular with a pointed tip, whereas, on the ventral side the shape of the denticulations (also of the same shape as those on the light colored ventral spots) is closer to being a trapezoid rather than a triangle. The stacking order of the denticulations on the ventral side differs also from that on the dorsal side. DDS are more compact and dense than those on the ventral side. The length, base width, and density of individual denticulations differ by location and color [10].

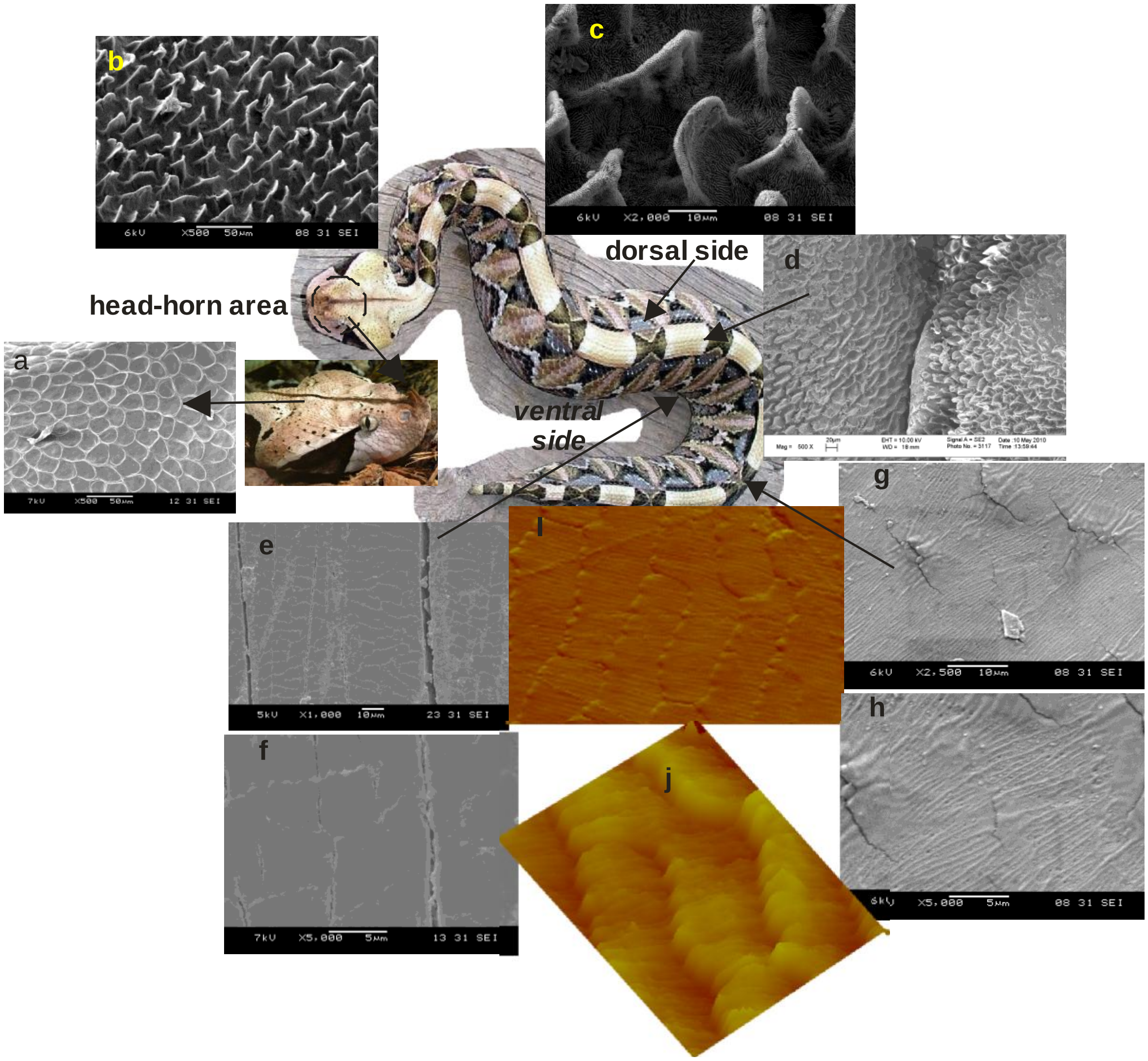

Figure 1: General appearance of the Bittis gabonica and SEM details of the three skin colors of the species: Dorsal skin light colored, Dorsal skin dark colored and the Ventral skin. All observations used a JEOL JSM-5510LV SEM acceleration voltage between 4 kv ≤ V ≤ 6 Kv).AFM images were obtained by using a Bruker Dimension Edge Fast Scan® AFM in topography mode

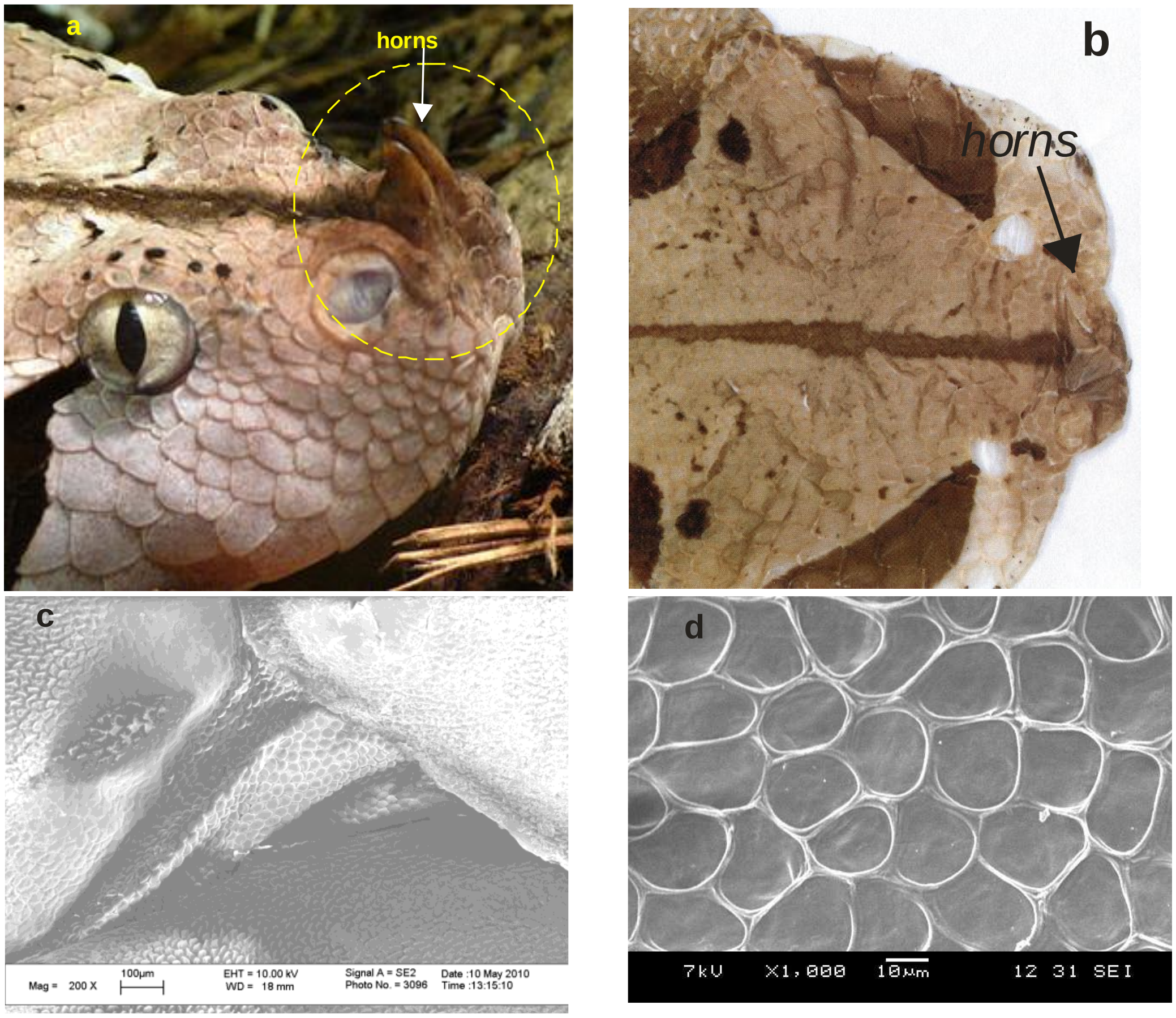

*Figure 2. General appearance of the Python Regius and SEM details of the three skin colors of the species: Dorsal skin light colored, Dorsal skin dark colored and the Ventral skin. All observations were performed on a JEOL JSM-5510LV SEM using an acceleration voltage that ranged between 4 kv ≤ V ≤ 6 Kv).*

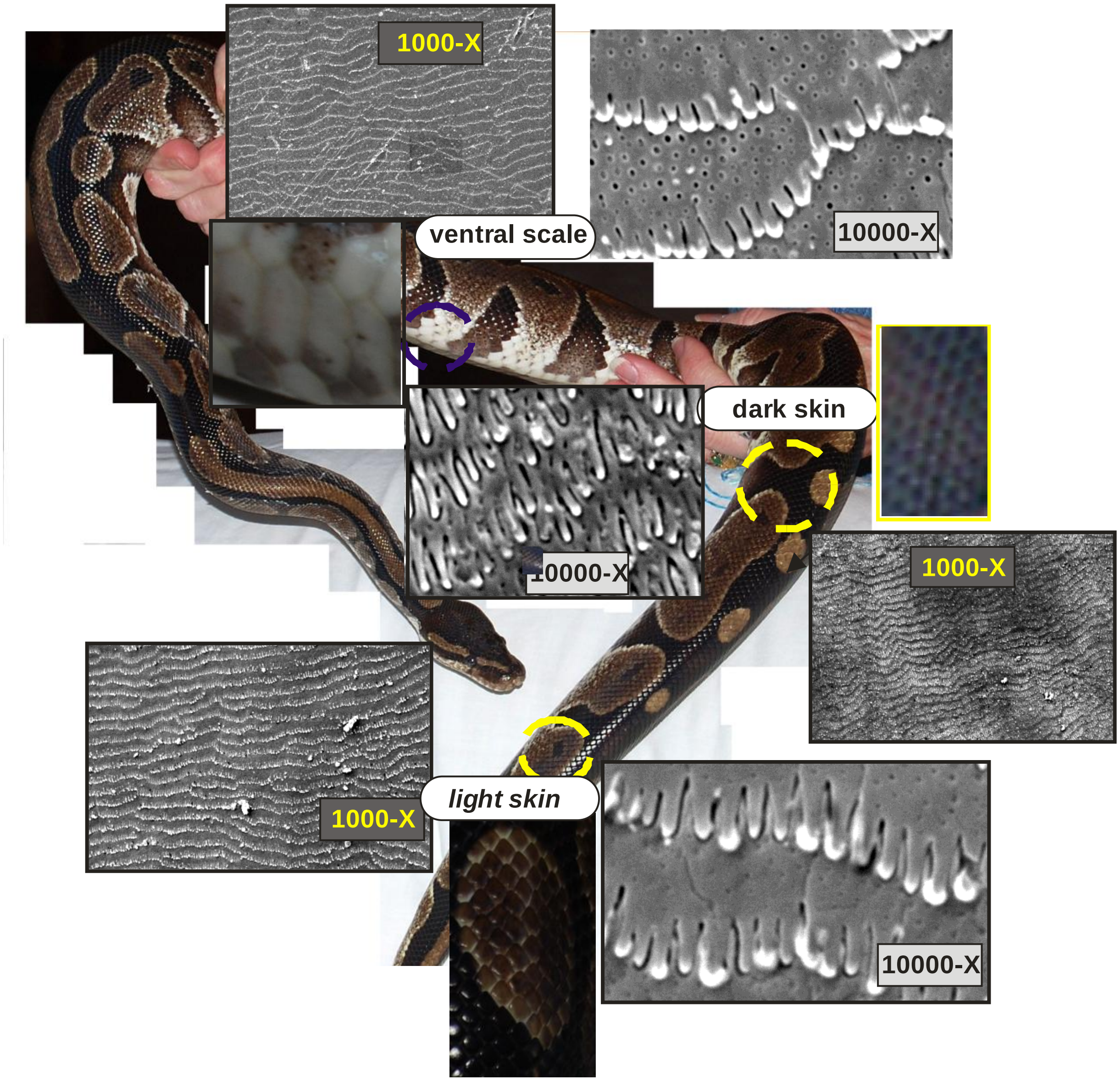

*Figure 3. General appearance of the Python Regius and SEM details of the three skin colors of the species: Dorsal skin light colored, Dorsal skin dark colored and the Ventral skin. All observations were performed on a JEOL JSM-5510LV SEM using an acceleration voltage that ranged between 4 kv ≤ V ≤ 6 Kv).*

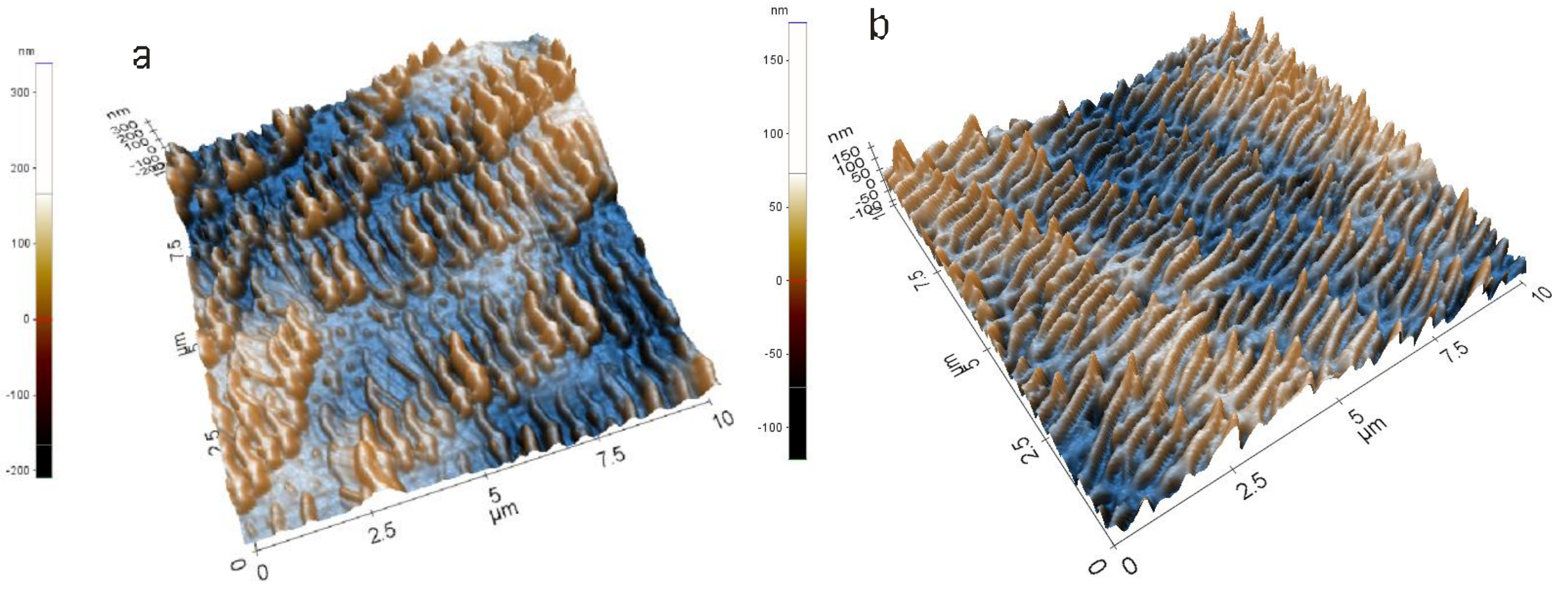

*Figure 4 AFM scans of 10 $\mu$m by 10 $\mu$m regions on the ventral (a) and dark-colored dorsal scales of the Python regius. Samples scanned using a Park Systems XE-150 cross-functional AFM in topography mode*

## 3. Methods

### 4.1.1 Skin treatment

The surface geometry of shed snake epidermis does not differ from that of a live animal [11, 12]. Therefore, the shed skin of snakes reflects the frictional response of the live animal. Furthermore, the shed skin reflects the metrological surface and textural features of the live animal. The current study used shed skin of two species: Python regius and Bitis Gabonica. The reason for such a choice is that while both snakes share the maximum length (see table 1) and mode of terrestrial locomotion (rectilinear locomotion, they differ greatly in weight. The Python regius weighs around 1.3 Kg whereas the Bitis Gabonica weighs around 8 Kg (see table 1). The difference in weight, while sharing the body length, forces several changes in body geometry and surface textural features. These manifest the functional adaptation of each species, which in turn facilitates the interpretation the observable differences in metrology of the surface and frictional behavior.

All information reported herein was extracted by testing skin obtained from five individual snakes of each species (i.e., five Pythons and five Bitis gabonica). Cleaning of the shed skin took place by soaking the received skin in distilled water kept at room temperature for 24 hours. Subsequent to soaking, the skin was allowed to unfold on cotton towels for two hours. This step allowed draining the excess water from the skin. Finally, we used compressed air to dry the skin. After final drying, the skin was wrapped in paper towels and stored in sealed plastic bags until experimentation.

*This work concerns the examination of skin obtained from shedding. In snakes, Skin shedding in snakes occurs naturally; as such, no injury to animals took place as result of in obtaining the examined materials.*

## 4. Surface texture metrology

### 4.1 Metrology of the Skin

To determine the metrological features of shed snakeskin, we identified thirty-two regions on the hide of each of the studied snakes. Each of the examined spots on the hide, of each species, comprised a section that is approximately 2.5 cm wide. Five WLI-interferograms were recorded for each of the dorsal and the ventral sides of the chosen region on the skin and these were analyzed to extract the surface texture parameters using a WYKO 3300 3D automated optical profiler system. To extract the surface parameters we used two software packages: Vision ®v. 3.6 and Mountains® v 6.0.

Figures 5 and 6 depict selected WLI-images for each of the snake species examined in the current work. Figure 5 (a through d) depicts two sets of images for the Bitis species. The first (figures 5-a and 5-b) pertain to two rectangular skin areas (approximately 1.25 mm by 1 mm) on the leading and the trailing halves of the ventral side of the snake (approximately at X/L= 0.3 and 0.7 respectively). Respective surface-height contours are color coded on the scale to the right of the images. Figures 5-c and 5-d, show WLI-images for the dorsal side. The image shown in Figure 5-c pertains to the junction between tow dorsal scales (i.e the so-called hinge region). The image of Figure 5-d depicts the surface in the middle of the scale. The WLI images shown in these figures present raw surface data in the sense that they contain the so-called large-scale roughness (curvature) of the skin superposed to the small-scale topography (or the deviation from main profile).

Figure 6 depicts WLI-images for the ventral (Figure 6-a, and 11-b) as well as the dorsal (figures 6-c and 6-d) skin spots for the Python species. The order of presentation follows that adapted in Figure 5. Thus, figure 6-a, depicts WLI-images of a ventral spot located on the leading half of the reptile and figure 6-b depicts the spot located on the trailing half of the ventral side of the reptile. The difference between the two instances is the scale of the studied spot. For the Bitis species, SEM images revealed that the denticulations are of a very small length and almost act as a serration within the surface. As such, it was judged that higher magnification will not affect the observation. For the Python, however, the shapes and distribution of the denticulations seem to be significant. This prompted us to use a higher magnification lens (X=50) to obtain the WLI-images. This resulted in the size for the examined ventral spots for the python being a rectangle of dimensions 125 μm by 100 μm, which is one tenth that of the spot size for the Bitis species shown in the previous figure. Note that the scale of presentation does not affect evaluating the statistical roughness parameters since these parameters are extracted from images of the same magnification. Figure 6-c presents the WLI-image of a square dorsal area of side 10 mm. This area contains several dorsal scales within the light colored portion of the python skin. General topography data are color coded to the right of the image. Figure 6-d, is an isolation of a single dorsal scale of those shown in figure 6-c. The isolated spot represents a square of approximate side 3.0 mm. Again, the images present both large scale and small-scale roughness data.

Figure 12 compares the values of selected surface roughness parameters extracted from figures 5 and 6. The values presented in the figure are distinct on two counts. First, they are an average of the particular parameter evaluated for all the locations examined on the body of each snake. Thus, in the presentation no distinction is made between the values of the local and the tailing halves of the reptile. Second, the figure presents the so-called Surface Parameters, which differ from the so-called profile parameters (typically designated by the letter R in metrology standards).

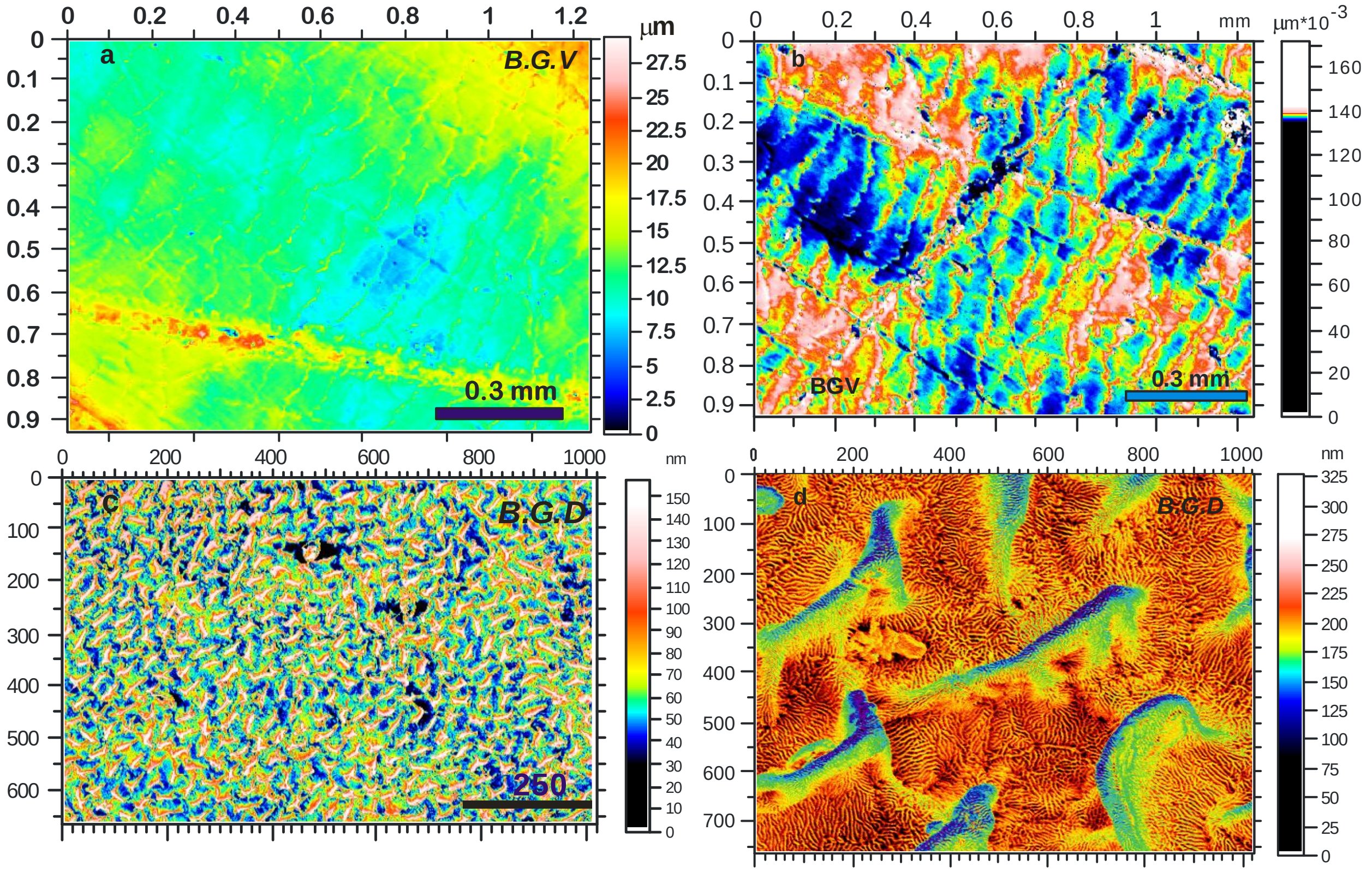

*Figure 5 WL-Interferograms of different spots on the skin of the Bitis gabonica; a and b pertain to ventral spots located within the leading and trailing halves respectively; c and d depict interferograms for two spots on the dorsal front half and on the middle section respectively*

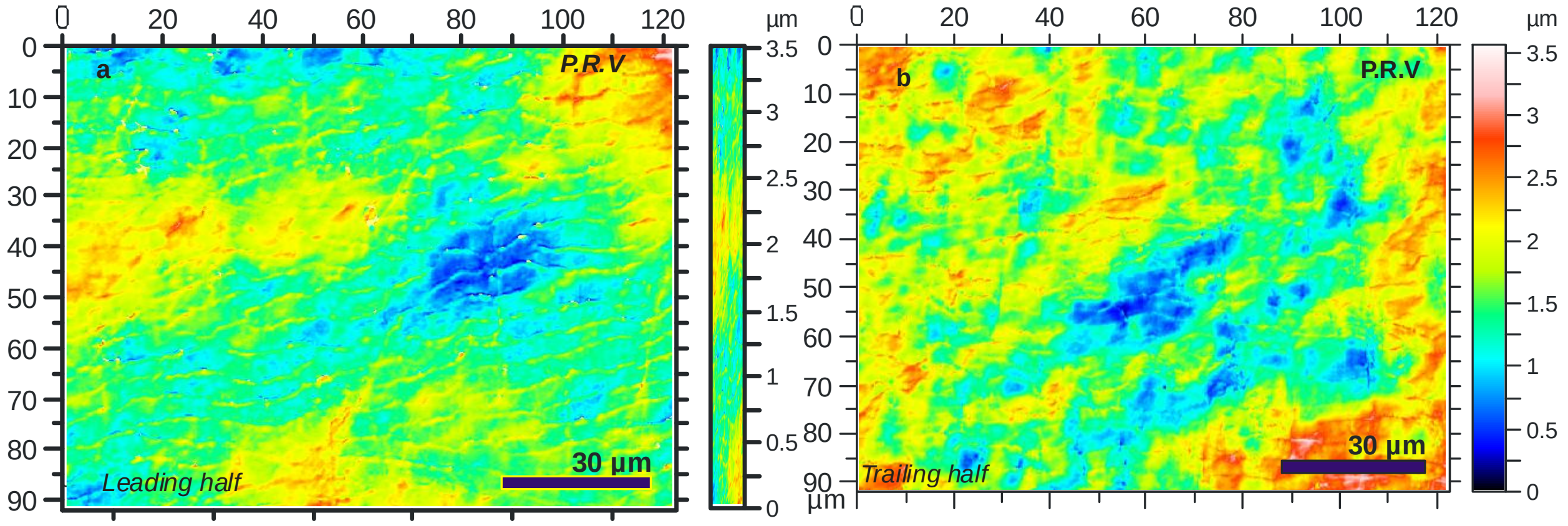

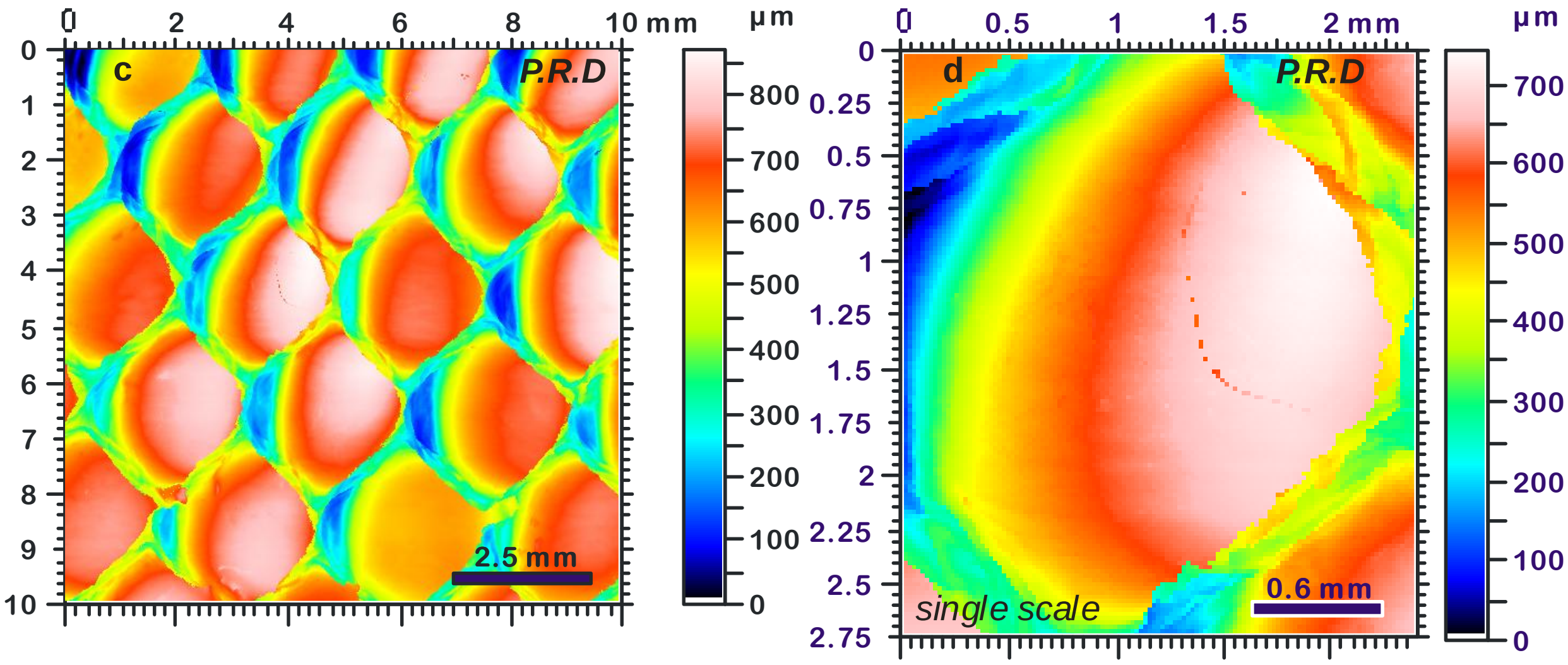

*Figure 6 WL-Interferograms of different spots on the skin of a Python regius; a and b ventral spots located within the leading and trailing halves of the snake respectively; c dorsal patch located within the light-colored skin, d topography of a single scale within the patch shown in c.*

The choice of the particular parameters to be presented stemmed from experience gained from previous work [13,14] which pointed at the importance of four roughness parameters in influencing frictional behavior of the skin. Each of the plots comprising figure 7 present the values for one of these parameters. Figure 7-a presents a plot of the average roughness parameter $S_a$. The figure depicts two sets of values: the $S_a$ for the Bitis species and the same parameter for the Python species. For each snake, the plot shows values for the ventral as well as the dorsal side of the skin. The lines atop of each of the bars represent the standard deviation of the measurements. Comparing values, one notes that protrusions within the Bitis ventral skin are higher than protrusions on the Python. The dorsal side manifests the same trend in roughness height. However, in this case the difference in the heights is clearly visible (roughness of the Bitis species almost twice that of the Python (the $S_a$ values are 0.75 μm and 0.4 μm respectively)). As expected, the RMS roughness parameter $S_q$ reflects a similar trend.
Figure 7-c is a plot of the kurtosis parameter $S_{ku}$ which represents the peakedness of the surface (with a value of around three indicating Gaussian distribution of heights). Again, the plot includes the data for the ventral and dorsal parts of the skin of both snakes. Examination of the values indicates that the values of the kurtosis, for both skin types, occupy a narrow band ($2.5 \leq S_{ku} \leq 3.75$). As such, the distribution of heights does not depart considerably from being Gaussian (or random). Focusing, however, on individual values reveals that the ventral side of the snakes is almost of equal kurtosis ($3.25 \leq S_{ku} \leq 3.7$ with the Bitis at the higher limit). This is slightly above the Gaussian limit. The spread among the $S_{ku}$ values for the dorsal side is wider (with the Python being at the higher limit).
From a statistical point of view, the skewness of the surface complements the physical description of the roughness. Figure 7-d presents the skewness data for the species. The figure compares the skewness parameter $S_{sk}$ for the dorsal and ventral sides of the examined skin. In contrast to the Kurtosis the Bitis ventral side displays opposite skewness to the python ($S_{sk}<0$ for the former and <0 for the latter). Meanwhile, for both snakes the dorsal skin is of positive skewness with that of the python slightly less than that of the Bitis.

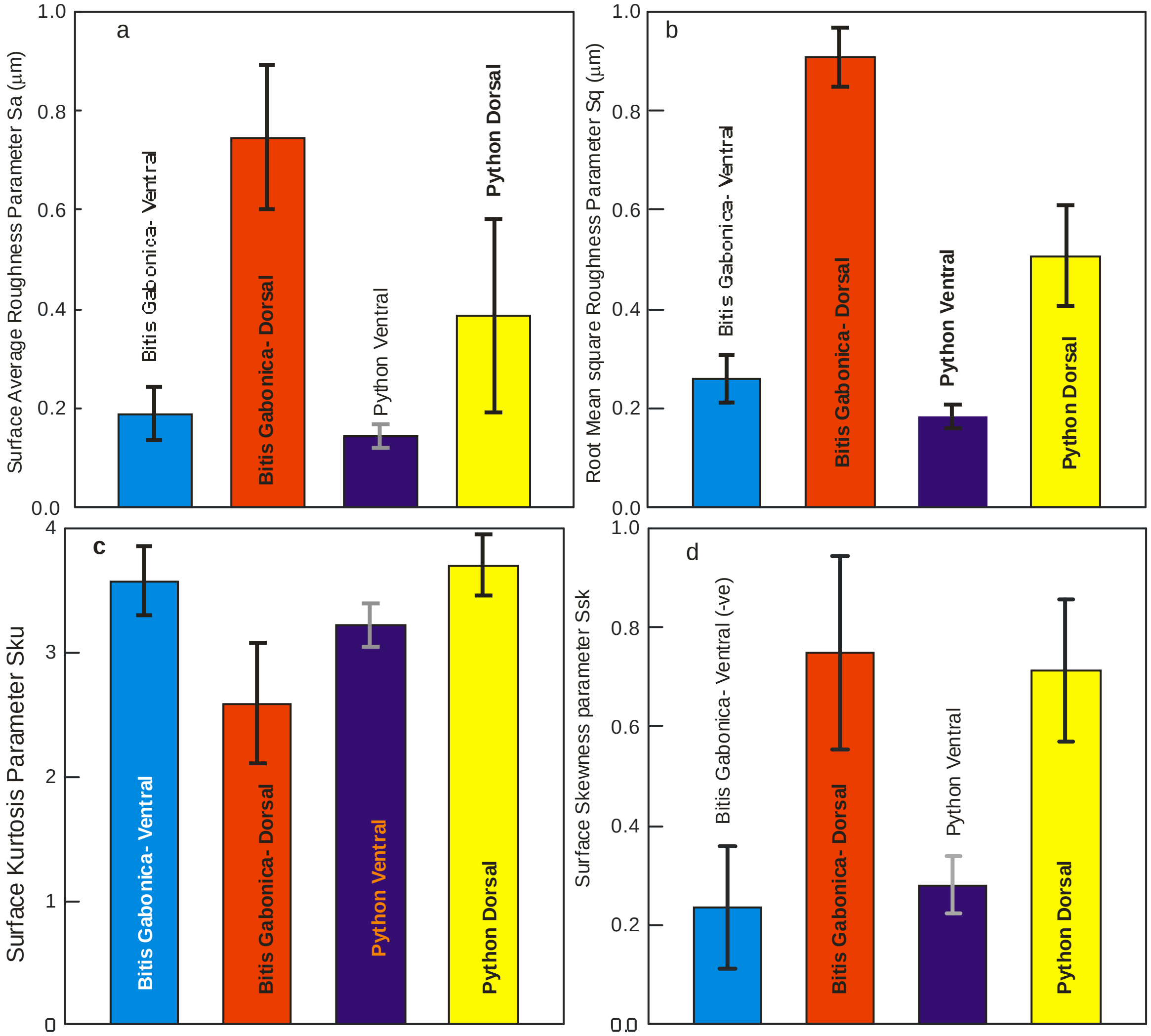

*Figure 7 Comparison between the surface roughness and aerial parameters of the examined reptilian surfaces; a- surface average roughness parameter, Sa, for ventral and dorsal sides of the Bitis and the Python species; b- root mean square roughness parameter Sq; c- surface kurtosis parameter Sku; and d- surface skewness parameter Ssk.*

$S_{sk}$ and $S_{ku}$ are the *Skewness* and *Kurtosis* of the 3D surface texture respectively. In a physical sense they manifest a topographical map of the heights of all measured points on the surface and the symmetry, or deviation, from an ideal Normal (i.e. Gaussian or random) distribution. The $S_{sk}$ parameter represents the degree of symmetry of the surface heights about the mean plane and its sign indicates whether peaks (i.e. $S_{sk}$>0) or valley structures ($S_{sk}$<0) dominate the surface. The parameter $S_{ku}$, on the other hand, informs of the presence of inordinately high peaks or deep valleys ($S_{ku}$ >3.00) or lack thereof ($S_{ku}$<3.00) within the makeup of the texture. An ideal random surface, i.e. for which surface heights distribution is Guassian, Skewness is zero and Kurtosis is equal to three. Surfaces of gradual varying roughness heights, free of extreme peaks or valley features, tend to have a Kurtosis less than three ($S_{ku} < 3.0$). The previous physical interpretation of the Skewness and Kurtosis parameters in the context of the surface layout allows conceptual visualization of the examined reptilian surfaces. For space limitation, however, we confine the analysis to the general layout of the ventral side.

The ventral surface of the Bitis manifests higher Kurtosis ($S_{ku} \approx 3.75$). Since the $S_{ku}$ is greater than three, we expect the presence of non-uniform peaks within the surface (which is natural due to the presence of the denticulations or fibrils (Figure 1)). However, since the surface displays negative skewness ($S_{sk} \approx -0.225$), we expect that depressions (or valleys) dominate the surface. Such a condition is possible if the spacing between two neighboring denticulations is relatively big than the width of the denticulation itself. On the other hand, the surface of the Python manifests a $S_{ku}$ that is also higher than three, $S_{ku} \approx 3.15$, which indicates the presence of protrusions above the plateau of the ventral scale. However, in contrast to the surface of the Bitis, the skewness of the Python surface is a positive quantity. Such a combination indicates, again keeping in mind the natural protrusion of the denticulations above the surface, that the spacing between the adjacent fibrils is smaller than that of the Bitis. That is, the packing of the fibrils within the ventral scale of the python is denser than that within the scales of the Bitis.

To verify this prediction, we re-examined the SEM images (figures 3 and 5) to determine the spacing between adjacent fibrils (denticulations) and the density of the fibrils within the respective ventral scale. Therefore, we evaluated the average spacing between fibrils and the ratio $A_{fib}/A_{VS}$ at each of the predetermined twelve positions on the body of each reptile (see figure 8). In all, for each spot fifty high magnification SEM images (X=10000) were examined. The results confirmed the shape of the reptilian surfaces as envisioned from roughness parameters. In particular, the "average" spacing between two neighboring fibrils was approximately 1.5 μm for the Bitis whereas, for the Python this spacing was around 0.5 μm. Moreover, the ratio $A_{fib}/A_{VS}$ was approximately 0.03 for the former and around 0.17 for the later.

### 5.1 Bearing curve analysis

When a normal force affects two complying rough surfaces, they establish contact through surface peaks. The first peaks to establish contact, are those of the longest height. Once in contact, the peaks deform bringing new pairs of opposite peaks, having an even smaller sum of heights, in contact. Thus, deformation leads to an increase in the number of peaks sustaining the load. Since peaks differ in height, the deformation of various peaks, within the same surface, will vary at any instant of time. As a result, only a fraction of the apparent area establishing contact will actually bear a load. The mechanistics of contact raise the question of predicting the area that actually supports a load beforehand. Conventionally, the answer comes in terms of a so-called Load Bearing Area Curve (Abbott-Firestone Load Curve (AFLC)) which is constructed from knowledge of surface topography. Study of the AFLC predicts the potential behavior of a given surface, of pre-characterized roughness, upon sliding and the likelihood of sustaining sliding-induced damage. Further analysis, of the AFLC yields predictive information about sliding performance formulated as standardized surface-functionality assessment metrics (the so-called $R_k$ family of parameters) [15-17].

Body girth in snakes is not uniform. It has a distribution along the Anterior Posterior Axis (AP-axis). Moreover, a snake does not establish contact with the substratum at all points of the ventral side. The irregular body shape of the reptiles results in local load-bearing requirements by the surface during locomotion. This is because of the resulting distribution of mass of the reptile and the forces required supporting, and propelling, the body segments. It follows that in order to understand the sense of load bearing performance in snakes there is a need to examine representative load bearing curves along the AP-axis.

In snakes, the variation in ventral scale lateral chord (i.e. the chord of an individual ventral scale along the lateral body axis, LL-RL-axis, of the reptile) is a good representative of the variation in body girth along the AP-axis. Therefore, following variations in the plot of that chord along the AP-body axis should point locations where local load bearing requirements change. In this work since the goal is to compare the sense of surface design for load bearing in snakes and cylinder liners, we elected to study the load bearing curves on the ventral scales of the two chosen snake species since the ventral sides are the main sliding surfaces for locomotion in snakes. To this end, we identified three zones on the hides of each of the examined snakes.

To identify each of these zones we first examined the variation of the length of the lateral chords of each species along the AP axis (figure 8-a). The figure depicts the variation in the local length of the local lateral chord along the body length of the reptile. The plotted values, denoted by $G/G_{max}$, represent the local length normalized by the maximum overall chord length. Based on the behavior of the plots we identified three zones in analogy of a cylinder liner. These are labeled I, II, and III in correspondence to the Top Dead Position (TDP), Medium Position (MP), and Bottom Dead Position (BDP), in cylinder liners. The criterion for identifying a particular position on the hide was to observe an appreciable variation in the slope in the $G/G_{max}$ curve. The resulting zones are depicted in figure 7-a. The extension B in the plot denotes the Bittis species while the extension P labels zones identified for the python species.

Figures 8-b through 8-d depict the extracted bearing curves at each of the pre-identified zones. Figure 8-b is aplot of the curves at the edge of zone I, figure 8-c presents the plots in the middle of zone II, and finally figure 7-d shows a plot of the curves at the edge of zone III. The figures imply that bearing curves for the Python are almost symmetrical more than that of the Bittis. The difference between curves of the two species is not apparent within the first two zones (zone I and zone II). However, the plots for zone III, located toward the tail of the species, manifest a visible difference between the species.

Notions of symmetry implied from graphical representation may be rather deceptive due to the scale of plots. It is of interest, therefore, to note the numerical values of the Peak Roughness height, $R_{pk}$, and the Valley Roughness Height, $R_{vk}$. If both parameters are equal, or close (i.e. $R_{pk} \approx R_{vk}$) in any one curve, then, the bearing curve for the particular zone is symmetric and vice versa. A symmetrical bearing curve indicates balanced interaction with the substratum upon sliding. This implies that upon contact, the peaks of the surface equally share the contact loading (weight in most cases). Absence of symmetry in the bearing curve, on the other hand, implies unbalanced contact whence contact forces and surface compliance will be a function of the higher of the two parameters ($R_{pk}$ or $R_{vk}$).

Table 2 presents a summary of the essential bearing curve values for both species. The table lists both the local and the average values of the $R_k$ parameters in addition to the average surface height Ra, for both species. The values of table 2 indicate that $R_{pk}$ and $R_{vk}$ differ by location. In each species, the Rpk value is almost equal for positions I and II. The value of that parameter within position III is almost double the values evaluated within the other two zones. The Rvk parameter however, does not reflect a similar trend. Rather, for the Bittis species the order of magnitude of $R_{vk}$ increases towards the posterior end of the reptile ($R_{vk}$-I < $R_{vk}$-II < $R_{vk}$-III), while for the Python this parameter is equal within zones II and III.

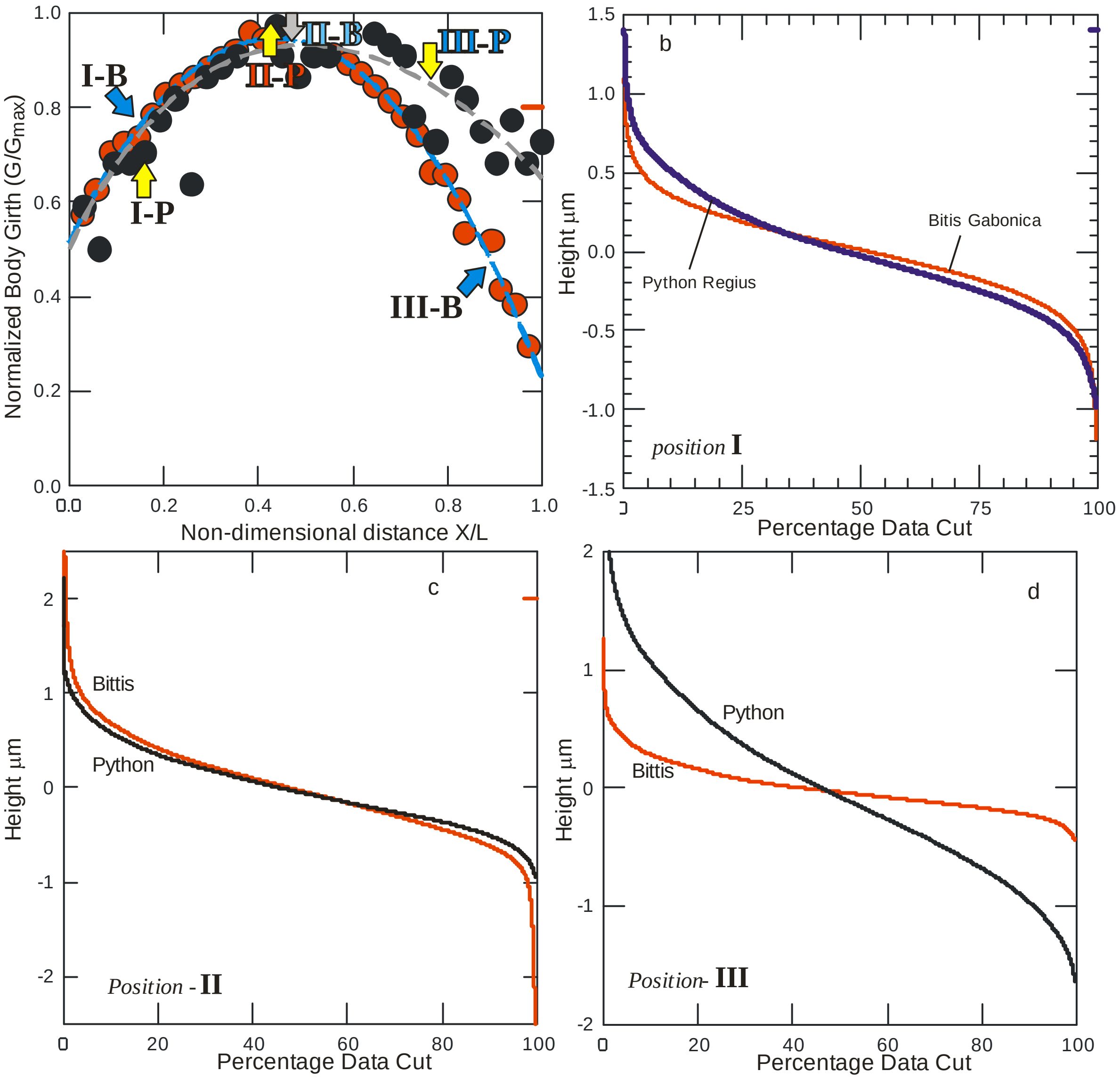

*Fugure 8 Comparison between the bearing curves for the snakeskin of the species examined in this work (P denotes Python and B denotes Bittis).*

Table-2 Surface and load bearing parameters for different regions within the skin as deduced from WLI-Interferrograms

| | $R_k$ | $R_{pk}$ | $R_{vk}$ | $S_a$ |
|---|---|---|---|---|
| Bittis gabonica | | | | |
| I-B | 0.295 | 0.322 | 0.145 | 0.16 |
| II-B | 0.445 | 0.31 | 0.631 | 0.326 |
| III-B | 0.994 | 0.546 | 0.854 | 0.364 |
| average | 0.578 | 0.393 | 0.543 | 0.283 |
| SD | 0.368 | 0.133 | 0.363 | 0.108 |
| Python regius | | | | |
| I-P | 1.203 | 0.44 | 0.788 | 0.393 |
| II-P | 1.211 | 0.463 | 0.407 | 0.374 |
| III-P | 1.987 | 0.934 | 0.449 | 0.625 |
| average | 1.467 | 0.612 | 0.548 | 0.464 |
| SD | 0.450 | 0.279 | 0.208 | 0.139 |

Further analysis of the bearing curve took place by evaluating the ratio of the values of the $R_k$ parameters to each other (table 3). The data imply that there is no location on the ventral side, of both snakes, that reflects complete symmetry of the bearing curve (although that for the Python species the "average" value of $R_{vk}/R_{pk}$ is almost unity).

Table 3 Ratio of the values of the $R_k$ parameters family for each of the examined species

| | $R_{pk}/R_k$ | $R_{vk}/R_k$ | $R_{vk}/R_{pk}$ |
|---|---|---|---|
| Bittis gabonica | | | |
| I-B | 1.092 | 0.492 | 0.450 |
| II-B | 0.697 | 1.418 | 2.035 |
| III-B | 0.549 | 0.859 | 1.564 |
| average | 0.779 | 0.923 | 1.350 |
| SD | 0.280 | 0.467 | 0.814 |
| Python regius | | | |
| I-P | 0.366 | 0.655 | 1.791 |
| II-P | 0.382 | 0.336 | 0.879 |
| III-P | 0.470 | 0.226 | 0.481 |
| average | 0.406 | 0.406 | 1.050 |
| SD | 0.056 | 0.223 | 0.672 |

It is of interest to examine the order of magnitude of the peak parameters in relation to the average height of the surface. At this point, it is beneficial to recall the nature of the topographical makeup of the ventral surface of both species as indicated from their kurtosis and skewness. The Bittis surface reflects negative skewness and a higher kurtosis, whereas the surface of the python reflects Guassian Kurtosis (around 3) and positive skewness. As such, the topographical layouts of the surfaces are almost opposite. The surface of the Bittis, being negatively skewed, tends to be more flattened and dominated y valleys. In contrast, the Python is less flat but dominated with peaks on the surface, which is more conducive to adhesion. The difference in masses between the two species ($M_B \approx 8.5$ Kg and $M_P \approx 1.5$ Kg), given the difference is surface structure, affects the stresses resulting while moving. In particular, for the Python, positive skewness will negatively affect the parts of the body where there is substantial surface-to-surface contact. Additionally, since peaks dominate the surface of the Python, contact stresses will intensify at the contacting peaks. This reflects on the requirements of load bearing and, in turn, leads to different peak and valley ratios on the bearing curves. To compare both surfaces there is a need to eliminate the effect of the difference in surface heights ($S_a$). To this end, an effective metric for comparison would be to relate the individual bearing ratios to the local average surface height ($S_a$) in each species. Table 4 presents a summary of the ratio of the bearing values to the local surface heights of both species.

Table 4: Ratio of the bearing parameters for the examined snakeskin to the average surface height.

| | $R_k/S_a$ | $R_{pk}/S_a$ | $R_{vk}/S_a$ |
|---|---|---|---|
| Bittis gabonica | | | |
| I-B | 1.844 | 2.013 | 0.906 |
| II-B | 1.365 | 0.951 | 1.936 |
| III-B | 2.731 | 1.500 | 2.346 |
| Average | 1.980 | 1.488 | 1.729 |
| SD | 0.693 | 0.531 | 0.742 |
| Python Regius | | | |
| I-P | 3.061 | 1.120 | 2.005 |
| II-P | 3.238 | 1.238 | 1.088 |
| III-P | 3.179 | 1.494 | 0.718 |
| average | 3.159 | 1.284 | 1.271 |
| SD | 0.090 | 0.192 | 0.662 |

The data show that the ratio $R_{vk}/S_a$ manifests contrasting trends in both species. In the Bittis the values of this ratio increase toward the posterior part of the body, whereas within the Python this trend is reversed. The ratio $R_{pk}/S_a$, on the other hand, increases toward the posterior parts of the body in the Python and displays a local minimum at the middle section of the Bittis (zone II). It is interesting to note that for snakes the modulus of elasticity increases toward the posterior of the body (the posterior parts are more rigid than the front parts) [5, 6]. As such, in the Python where adhesion is most likely to dominate, because of positive skewness, the rigid zones of the body have higher average height and lower reduced valley depth. Such a trend is reversed in the Bittis species. The reduced peak height value ($R_{pk}$), meanwhile, increases with the rigidity of the surface in both species. The core roughness depth $R_{pk}$, however, remains unaffected by rigidity in the Python.

## 5. Correlation to honed surfaces

The principal idea of honing is to enhance the wear and sliding performance of the cylinder liner by imposing a geometrical pattern on the external layer of the liner. The texturing of a honed surface entails two components: a raised protrusion called "*plateau*" and an entrenched component known as a "*groove*". The grooves, in principal, retain lubrication oil during piston sliding and thereby they replenish the lubrication film in subsequent sliding cycles. The plateau, provides raised cushions (islands), that the piston will contact upon sliding, so that the total contact area is reduced. A system of such nature is multi-scale by default. Thus, the basic metrological characteristics of a honed surface will depend on the scale of observation.

One of the major requirements for an optimal honed surface is connectedness. Through perfect connectedness between all unit-texture features, high lubrication quality, and economical lubricant consumption, takes place.

To date there are no standardized values of the relevant surface parameters that ensure superior performance of a honed cylinder liner. Instead, each manufacturer has an "in-house" set of recommended values that surface parameters should fall within for quality performance. To this end, comparing the metrological features of snakeskin to recommended values is qualitative in essence. Never the less, comparing bearing curve parameters of both skin and cylinder liners yield valuable information about both surfaces. In this section, we perform this comparison using two sets of data. The first set represents the bearing curve parameters of unused commercial cylinder liners obtained from an auto manufacturer. The second set of data meanwhile represents the surface parameter recommendations of a commercial cylinder liner supplier [18-20].

### 5.1 Comparison to Commercial cylinder liner

To compare skin and a commercially honed surface, segments of a commercial unused engine prepared by a sequence of honing processes (see appendix-I). Figure 9 depicts three representative segments of the examined liners. The figure details the liner surface at three positions, TDP, MP, and BDP respectively. For each of the segments, we present SEM images, WLI-scans, and the relevant bearing curve. Table 5 provides a summary of the $R_k$ family of parameters extracted from WLI-scans for the particular locations and the ratios of the bearing curve parameters to the average surface height $S_a$. The ratios provided in the table reveals some similarities between the liner surface and that of the snakes (table 4).

The liner ratio $R_k/S_a$, for example, is in the same order of magnitude in all positions (TDP, MP, and BDP). The python data reflect a similar trend, whereas the Bittis has a minimum value at the middle position (position II-B). The Python and the cylinder liner both share the same trend of the ratio $R_{pk}/S_a$. This ratio increases from top to bottom within the cylinder liner and from anterior to posterior ends within the Python. Similarly, the ratio $R_{vk}/S_a$ meanwhile decreases from top to bottom within the liner and toward the posterior end in the Python. The Bittis, however, contrasts this trend as this ratio increases toward the posterior end of the reptile. The data reflect acceptable similarity in bearing curve parameters of the liner and the Python more than the parameters of the Bittis. As such, from this point on, we will confine further comparison with honed surfaces to analysis of Python data.

The basic textural pattern in a honed surface is a parallelogram (figure 10) whereas, within the ventral scale of the python, is a high aspect ratio trapezoid (figure 4-a). The aspect ratio of the honed surface is constant along the length of the liner while in a python this ratio changes along the AP-axis [21]. Common between the two surfaces is the elevation of the element of texture above the general plateau of the supporting substrate. The elevation above the substrate in the liner is, again constant along the length of the cylinder. One apparent, but deceptive, difference between the two surfaces is the orientation of the unit textural elements on the surface of the liner, for the technological surface, and on the scale for the snake surfaces. In snakeskin, the fibrils generally point at the posterior of the reptile. However, the placement of the fibrils on the plateau of the scale follows a wave-like arrangement (see figures 2-4). The orientation of the fibrils, however, with the respect to the AP-axis follows from the nature of motion of the snake and the manner it contacts the substratum. The body of the snake is flexible and does not contact the substratum at all points upon moving. Due to the irregular terrain that the snake encounters, a definite orientation of the fibrils along the AP-axis is not feasible since fibrils modify contact conditions. To maintain efficiency and versatility textural elements have to be oriented along almost all possible contact points of the body. A honed surface, such as that of the cylinder liner, does not encounter surfaces as irregular as natural substrates. Moreover, the placement the cylinders within an engine block confines sliding of the contacting surfaces, on the macro scale, to a motion along a single axis. As such, the need for periodic alteration of the orientation of the textural elements does not arise.

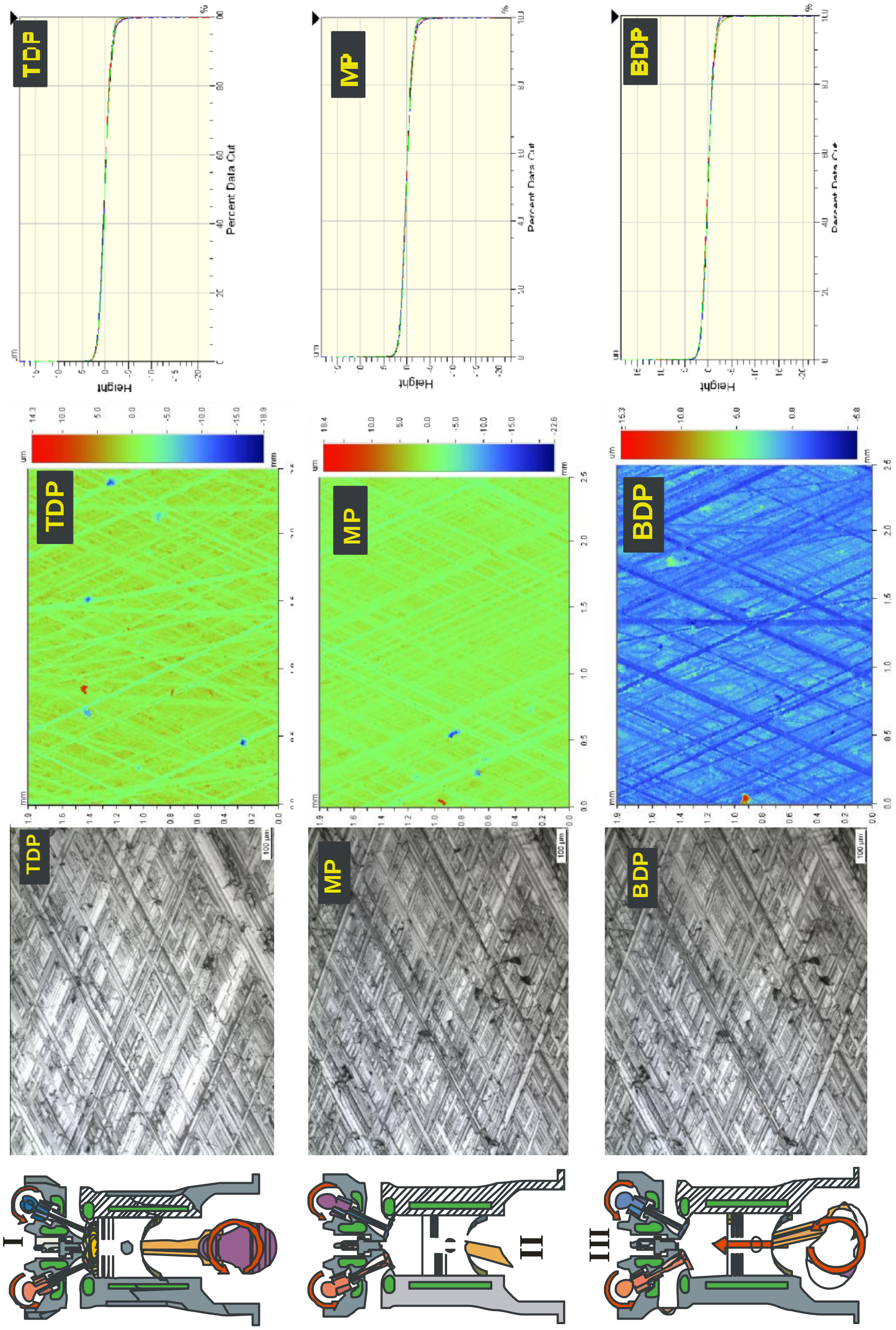

*Figure 9 Details of the control honed liner surface. The figure depicts SEM images, WLI images, and respective load bearing curves at three spots within the liner. The data pertains to the Top Dead Position (TDP), Medium Point (MP), and the Bottom Dead Position (BTP) respectively. Conditions and manufacturing stepsfor the surface are summarized in Appendix-I.*

Table 5 Values of bearing and surface parameters of the honed segments shown in figure 8.

| | $R_k$ | $R_{pk}$ | $R_{vk}$ | $S_a$ |
|---|---|---|---|---|
| TDP | 3.1 | 1.21 | 1.71 | 0.978 |
| MP | 2.697 | 1.17 | 1.44 | 0.854 |
| BDP | 2.813 | 1.34 | 0.86 | 0.867 |
| | $R_k/S_a$ | $R_{pk}/S_a$ | $R_{vk}/S_a$ | |
| TDP | 3.16 | 1.23 | 1.75 | |
| MP | 3.15 | 1.37 | 1.68 | |
| BDP | 3.24 | 1.54 | 0.99 | |

Despite the difference in orientation of the elements of texture in both surface types, the spacing between the textural features, within each surface, is almost uniform. This arrangement provides for enhanced oil film distribution within the liner surfaces and for quasi-uniform local contact for the snake surface. An additional feature common to both surface types is the rather shallow depth of the separation between the elements of texture. Shallowness enhances wear resistance in honed surfaces. To the knowledge of the authors, the advantages of this geometry on locomotion or friction for reptiles have not been examined. However, accepting the logic of construction of honed surfaces, one may predict that combat of wear is one of the advantages.
Superficial depth of the spacing offers another advantage with respect to trapping, dust and erosive contaminant particles. Trapping of particles enhances resistance to abrasion by such particles, which agin contributes to the structural integrity of surfaces in each case. In essence, it seems that general paradigm of texture engineering in both surfaces follows along the same broad lines. It to be noted that construction of each of the class of surfaces examined evolved along entirely different paths. This supports the feasibility of integrating some of the geometrical features of bio-surfaces into technological surfaces.
Upon sliding, due to the nature of contact and wear the roughness will change. Such a change affects the bearing characteristics of the surface of the liner. In the Python surface, the bearing parameters or precisely their ratios do not vary by scale. This possibly points at the fractal nature of the Python surface. To verify this preposition we evaluated the fractal dimension of selected locations within the Python ventral skin and the analogous position son the cylinder liner from a series of SEM images. The potential wide range of magnifications that the SEM imagery offers facilitates testing the fractal nature of the respective surfaces.
The fractal dimension of the surfaces was computed using the cube counting method. Three magnifications were chosen, 100 X, 250 X, and 5000 X. Figure 11 presents a comparative plot of the resulting fractal dimension of the skin and the liner surfaces.
Fractals efficiently describe the geometry of regular shapes that are not amenable to conventional description through Euclidean geometry. In particular, fractals describe how much space an irregular object occupies [22]. The structure of the fractal number is indicative of the nature of the object described. The most significant digit of the parameter, D, stands for the topological dimension of the object, whereas the second part, the fractional part that varies from 0.0 to 0.999, is a so-called fractal increment. The higher the fractal dimension, the more space the curve occupies. That is, the fractal dimension indicates how densely a phenomenon occupies the space in which it is located. Thus, the fractal dimension does not necessarily describe the micro texture of a surface. Rather, it reflects the dependence of the surface on its roughness to occupy a three dimensional space. By focusing on the growth of form, and its evolution in space, fractal description of a surface shifts the focus from devising Euclidean geometry to describing the evolution of its growth into its prescribed space. That is the fractal description of a surface deals with the evolution of the process of surface growth into the space rather than finding its boundaries with respect to an origin. As such, a fractal frame of reference projects differences within the micro-textural elements, comprising the surface, in a view that unmasks their common phenomenal traits. This accentuates the correspondence between behavioral patterns, of apparently dissimilar Euclidean form, especially when responding to tribological phenomena (e.g. contact forces or diffusion through surface topography).

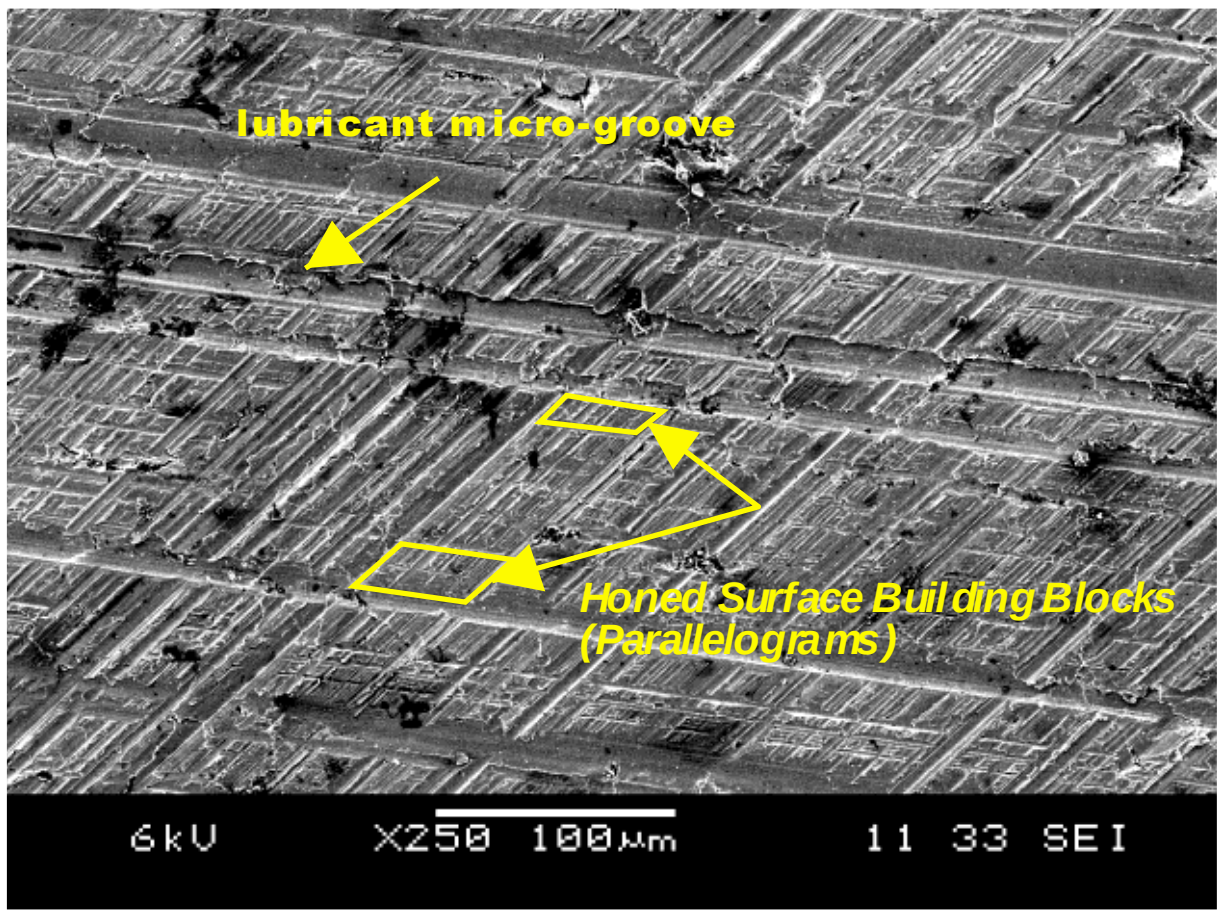

*Figure 10: SEM-Micrographs of a cylinder liner sample used for comparison with the Python skin, X-250.*

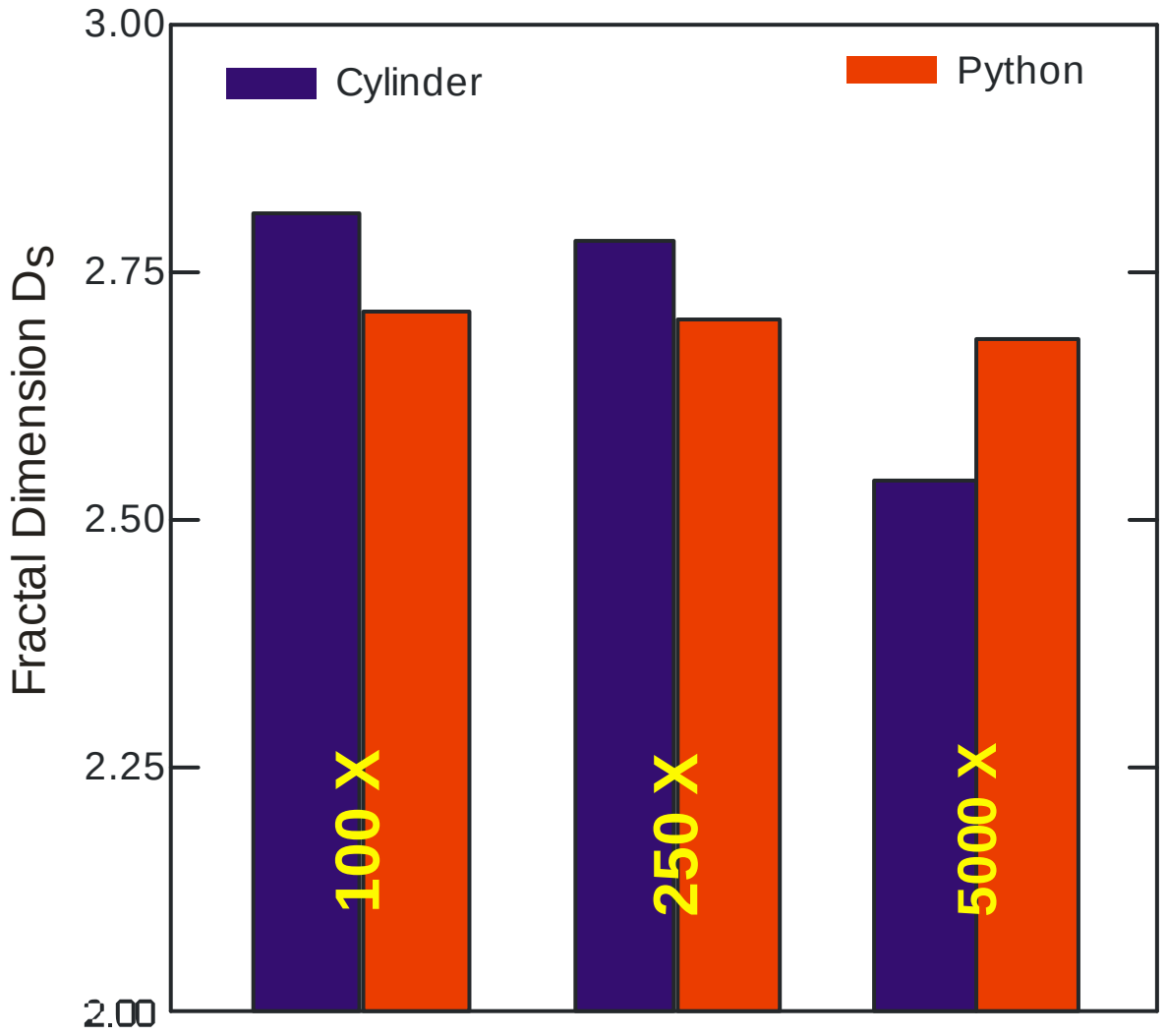

*Figure 11 Variation in the fractal dimension, D, of the Python skin ventral side and the cylinder liner with scale of magnification. The fractal dimension was computed from analysis of SEM images taken at three magnifications (100 X, 250 X, and 5000 X).*

Initial study of figure 10 reveals a high fractal dimension of the Python skin and the cylinder liner (D> 2.5). This implies that both surfaces depend on roughness to fill their space. It also implies that for both surfaces the properties of the roughness dominate the behavior more than the bulk of the base materials.
A major difference between the two surfaces is the variation in the respective fractal dimension with magnification. The fractal dimension for the cylinder liner decreases with magnification (figure 11). For the python surface, however, the fractal dimension is almost invariant with magnification. The fractal dimension for the reptilian surface, moreover, is smaller than that of the cylinder liner. Invariance of the fractal dimension of the reptilian surface points at potential isotropy of the surface. It also may imply that the variation in the parameters of the bearing curve for the Python show uniform variation with respect to scale of observation. To verify this observation, we extracted the load bearing parameters of the Python and the cylinder liner from the SEM pictures used to extract the fractal dimension. Figure 12 (a and b) plots the results of the analysis. The figure depicts metrological surface parameters calculated at two scales of observation: X-250 and X-5000.
For *qualitative* analysis, we recall the role of the parameters $R_{pk}$, $R_{vk}$, and $R_k$ in ranking the sliding performance of a given surface. $R_{pk}$ indicates the height of the surface that wears away during the run-in period. Thus, it is preferred to be at a minimum. $R_k$ indicates the portion of the surface that supports the load after running. It is preferred to be of a relatively higher value. $R_{vk}$ meanwhile indicates the lowest portion of the surface that will retain lubricant (for minimum oil consumption is required to be relatively high). Ideally, the maximum value of the ratio $R_{pk}/R_k$ should be around 0.5, $R_{vk}/R_k$ should be less than unity (at or around 0.5), and $R_{vk}/R_{pk}$ should be around unity. Keeping those estimates in mind, we note that the Python skin is slightly off the target limits at large scale (X-250). However, when considering smaller scale (or the higher magnification X-5000) the surface reflects an opposite trend. Interestingly while the ratio $R_{pk}/R_k$ is almost within qualitative limits for the honed surface at large scale (X-250), it significantly departs from that limit at smaller scale. The opposite is noted for the ratio $R_{vk}/R_{pk}$. Such an observation prompted the calculation of the percentage of change in the three examined geometrical proportions as a function of scale of observation. Figure 13 plots the results of computing the percentage variation in bearing parameters at two different magnification scales. This ratio, we envision, is a conservative prediction of the evolution of wear of the surface. This is because the asperities of shorter heights above the surface will come to support the contacting load upon wearing of the asperities of higher elevation. As such, computing the ratio of change would indicate the evolution of bearing characteristics as a function of recession of the surface. The skin of the Python has uniform and minimal variation in the surface geometrical proportions than the liner surface (for which the ratio $R_{pk}/R_k$ shows the largest variation). Such a wide variation may affect performance of the honed surface at very small scale through affecting the connectedness of the microgrooves and thereby the unobstructed flow of the lubricant while sliding.

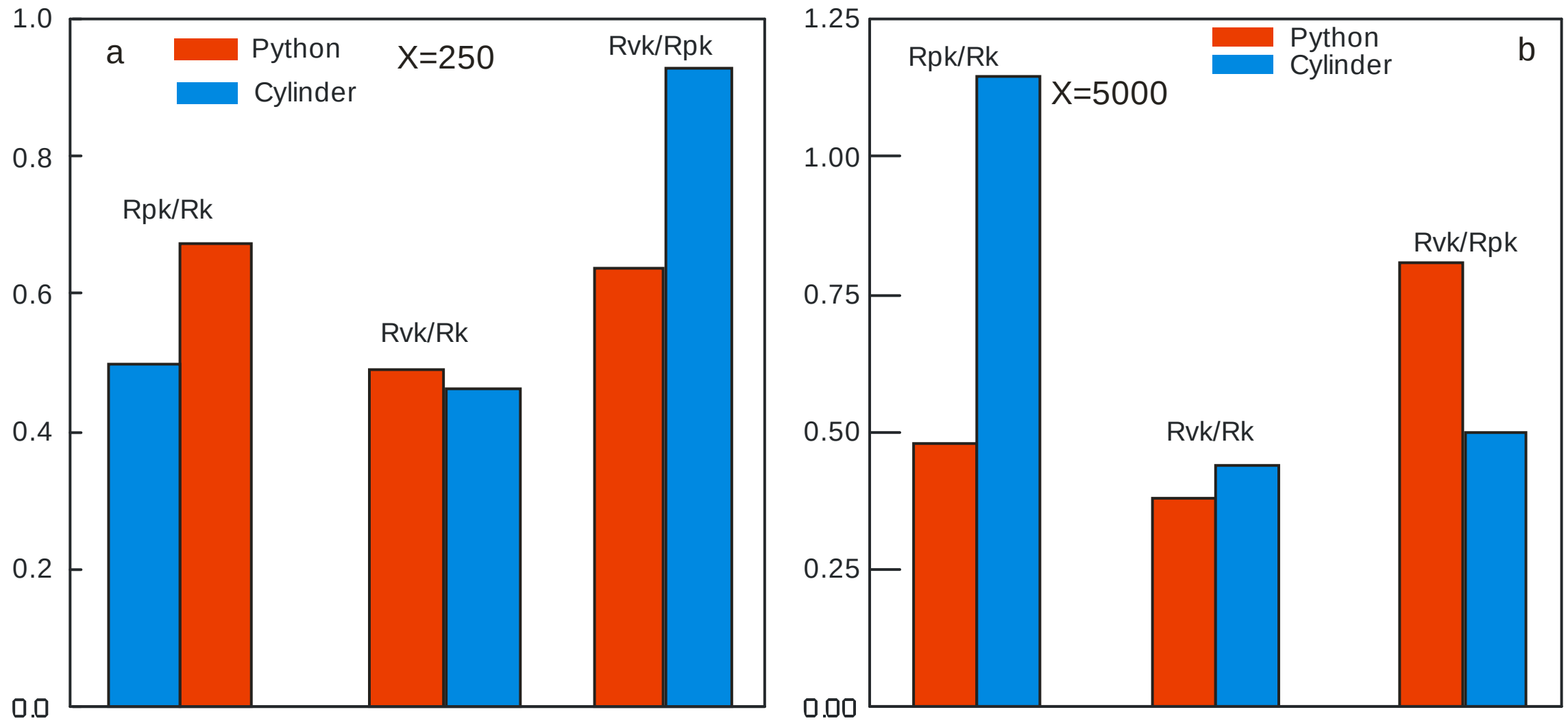

*Figure 12: Comparison between the geometrical and metrological proportions of the Python skin surface and those of a Plateau honed cylinder liner at two scales of SEM observation X-250 and X-5000.*

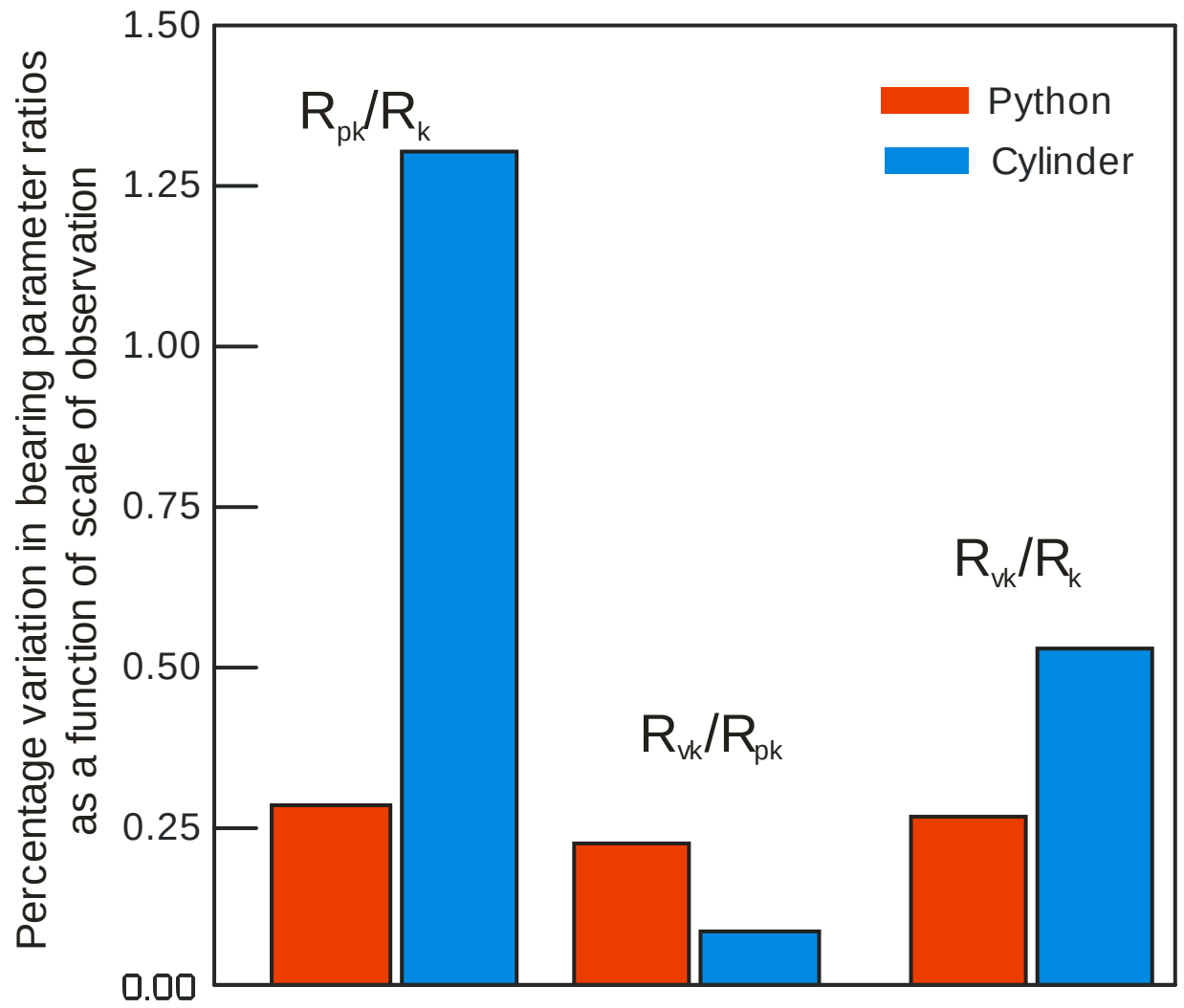

*Figure 13: Absolute value of percentage variations in surface proportions as a function of scale of observation.*

## 5.2 Comparison to recommended texture values

The second set of analysis performed in this work compares bearing characteristics of the two snake species to recommended and actual surface finish values in commercial applications [18-20]. The bearing characteristics in this section are represented by the ratio of the various peak family parameters ($R_{pk}$, $R_k$, and $R_{vk}$). Appendix II provides a summary of the raw extracted data and manufacturing steps used in producing the surfaces.

Figure 14 (a and b) provides a graphic comparison between the average bearing parameter ratios for the snakeskin and the extracted ratios. The figure lists the values the recommended values (circles), and the comparative R parameter ratio for the snakes (hexagonal and rhombic symbols). The figure indicates that the bearing ratios for snakeskin are at the lower end of the set of recommended values (very close to the values of the precision fine finished surfaces).

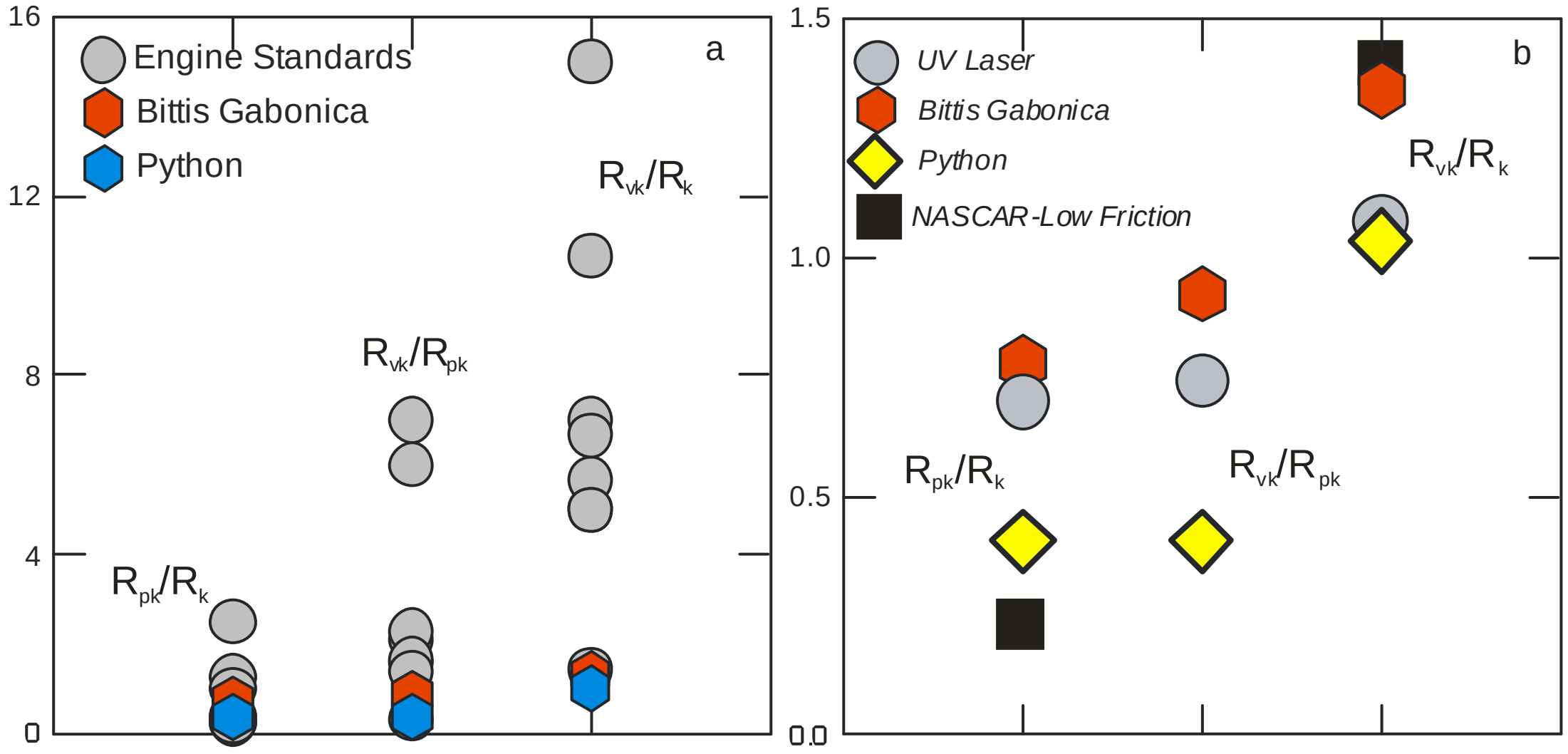

*Figure 14 Comparison of the ratios of the $R_k$ family of parameters and some recommended values for surface finish for commercial cylinder liners (see appendix II). Figure 14-a depicts a comparison with the entire set of recommended values. Figure 14-b provides a comparison with the ratio of parameters for precision UV Laser finished surfaces and recommended values for minimized friction for NASCA engines.*

Figure 14-b compares the bearing ratios of the two snake species to bearing ratios of two precision finished surfaces. The first (grey circles) is a fine precision surface finished using UV Laser (see appendix II). The second value represents the finishing values recommended for low friction operation for NASCAR racing engines (square symbols). It is noted that the values for the snakes enclose those for the technological surfaces when comparing the ratio $R_{pk}/R_k$. The value of this ratio for the Bittis (hexagonal symbol) is higher than that for the laser-finished surface, and higher than that for the python. The value for that ratio for the python (diamond symbol) is slightly higher than that for the reduced friction recommendation. The python has the lowest value for the two other ratios (although that for the $R_{vk}/R_k$ the python and the laser-finished surface are almost equal). The value of $R_{vk}/R_k$ implies that the snake surface is equipped for lubrication ($R_{vk}/R_k \approx 1.0$). This may relate to the habitat of the species and their respective body masses. Both examined species move by using rectilinear locomotion. However, the Bittis is almost five times heavier than the python, whereas the length of both snakes is almost equal. Both species meanwhile co-exist within the general geographic area [8]. The Bittis species tends to be within the rainforest in high rainfall areas. The python meanwhile tends to inhabit grassland and savannas. For the heavier snake, moving in a rainy area provides water lubrication to counteract the high pressure exerted by the higher body mass. Two requirements are essential for optimized locomotion in this case: lower friction and less protrusion of the surface peak to reduce skin abrasion. The first requirement is met through high ventral scale aspect ratio in the lateral direction (consequently less friction and larger nominal bearing area [23]). Locally, however, $R_{vk}$ is considerably deep to preserve any friction reducing agent (moisture or slurry). For the python, body mass is smaller so the friction requirements are not as strict in comparison to those for the Bittis. The valley parameter $R_{vk}$ is almost one-half of the peak parameter $R_{pk}$. The peak parameter meanwhile is a fraction of the bearing portion of the surface. The peak parameter being higher compensates for the positive skewness of the skin, which favors adhesion and indicates that peaks dominate the surface. Such a condition promotes gripping by the surface, which subjects the peaks to wear. It is interesting, in that context, that the peak parameter of the python surface, despite being relatively high, is close to the value recommended for reduced friction in highly loaded racing cylinder liners. The valley parameter, meanwhile, matches the recommendation for minimal wear.

## Conclusions

In this work, we presented a comparison between the load bearing characteristics of snakeskin and technological honed surfaces. The study examined two snake species, Bittis gabonica and Python regius.

There are several similarities between the skin surface and the honed surfaces. The unit building block of the surface features in each case is a quadrilateral (parallelogram for the honed surface and a high aspect ratio trapezoid in a snake surface). In snakes, the aspect ratio of the surface features varies by location while it is almost constant in the honed surface. Spacing between the elements of texture within each of the surfaces is uniform.

The shape of the load bearing curve for reptilian skin is quasi symmetric on average (although locally this symmetry may be violated).

The results indicate that the skin of the python is closer in proportion to the examined cylinder liner surface and falls within the recommended critical surface finish values recommended by commercial liner manufacturers.

Reptilian surfaces seem to be of fractal nature. Calculation of the fractal dimension implied that the roughness controls the behavior of the surface rather than the bulk of the material. This reflects on the hierarchical nature of the biological surface. The examined surfaces manifested a uniform decline in the bearing parameter ratios.

Finally, bearing parameter ratios for the reptilian surfaces manifested a high degree of specialization in the sense that the bearing parameters meet the local requirements for structural integrity. The bearing parameters enhance the wear resistance and counteract the consequences of local skewness on adhesion. It is of interest that the logic of texture construction of the reptilian and technological surfaces bear several similarities despite that each have evolved along entirely different paths.

Appendix -I

Table I-1: Operation parameters implemented in the sequence of operations used in preparing the examined honed surfaces shown in figure 9

| | | |
|---|---|---|
| | ABRASIVE | |
| | Grain type | Diamond-bond bronze) |
| | Number of stones | 8 |
| | EXPANSION TYPE | Mechanical |
| | HONING TIME (sec) | 30 |
| | ROTATION SPEED (tr/min) | 280 |
| | AXIAL SPEED (m/min) | 22 |
| | EXPANSION SPEED 1 (µm/s) | 4 |
| | EXPANSION SPEED 2 (µm/s) | 35 |
| FINISH HONING | ABRASIVE | IAS |
| | Grain type | Silicon Carbide (bond: vitreous) |
| | Number of stones | 8 |
| | EXPANSION TYPE | EMZ |
| | HONING TIME (sec) | 25 |
| | ROTATION SPEED (tr/min) | 180 |
| | AXIAL SPEED (m/min) | 22,2 |
| | HONING ANGLE | 50,1 |
| | EXPANSION SPEED 1 (µm/s) | 8 |
| | EXPANSION SPEED 2 (µm/s) | 1.5 |
| PLATEAU HONING | ABRASIVE REFERENCE | SC |
| | Grain type | Silicon carbide |
| | Number of stones | 4 |
| | EXPANSION TYPE | Hydraulic |
| | HONING TIME (sec) | 5 sec |
| | ROTATION SPEED (tr/min) | 180 |
| | AXIAL SPEED (m/min) | 22,2 |
| | HONING ANGLE | 50,1 |
| | Pressure (bars) | 8 |
| | Number of strokes | 4 |
| COOLANT | *Mineral Oil* | |
| WORKPIECE | Lamellar Gray cast iron, 3.1-3.5% carbon, 187-235 Brinell hardness | |

Appendix II

Table II-1 Summary recommended values for surface finish of highly loaded cylinder liners.

| Honing types | Inductive hardening | Laser hardening | Honing lapping | Plateau honing | Slide honing | Helical honing | UV laser treatment |
|---|---|---|---|---|---|---|---|
| Structured area [mm] | 20 | 20 | entire length of cylinder | entire length of cylinder | entire length of cylinder | entire length of cylinder | entire length of cylinder |
| **Work steps** | | | | | | | |
| Step 1 | Inductive hardening | Laser hardening | | | | | |
| **Honing Type** | | | | | | | |
| Step 2 | Coaxial | Coaxial | Pre | Pre | Pre | Pre- | Pre- |
| Step 3 | Honing | Honing | Lapping | Honing | Honing | Honing | Honing |
| Step 4 | Finish | Finish | Finish | Finish | Finish | Finish | Finish |
| Step 5 | | | | | | | UV lasering |
| **Surface parameters** | | | | | | | |
| angle [°] | 33 | 33 | 35 | 33 | 40 - 60 | 140 - 150 | 45 |
| $R_z$ [µm] | 7 | 7 | 7.5 | 7 | 2.8 | 2.9 | 4.99 |
| $R_{pk}$ [µm] | 0.4 | 0.4 | 0.8 | 0.4 | 0.08 | 0.1 | 0.99 |
| $R_k$ [µm] | 1.5 | 1.5 | 2 | 1.5 | 0.33 | 0.5 | 1.42 |
| $R_{vk}$ [µm] | 2 | 2 | 1.3 | 2 | 1.03 | 1.8 | 1.05 |

Table II-2 Recommended values for Honned Surfaces

| $R_k$ (µm) | | $R_{pk}$ (µm) | | $R_{vk}$ (µm) | | $R_a$ (µm) | |
|---|---|---|---|---|---|---|---|
| General Applications | | | | | | | |
| min | max | min | max | min | max | min | max |
| 25 | 35 | 8 | 12 | 40 | 50 | 15 | 20 |
| Reduced Friction Short Life (Racing NASCAR) | | | | | | | |
| 12 | 18 | 3 | 5 | 20 | 25 | 7 | 11 |

| Rpk/Rk | Rvk/Rk | | | | Rvk/Rpk | |
|---|---|---|---|---|---|---|
| General Applications | | | | | | |
| max | min | max | min | max | min | max |
| 2.500 | 0.320 | 0.343 | 1.600 | 1.429 | 5.000 | 4.167 |
| Reduced Friction Short Life (Racing NASCAR) | | | | | | |
| 2.273 | 0.250 | 0.278 | 1.667 | 1.389 | 6.667 | 5.000 |